\documentclass[fleqn,usenatbib]{mnras}

\usepackage{threeparttable} 
\usepackage{newtxtext,newtxmath}
\usepackage{multirow}
\usepackage{anyfontsize}

\usepackage[T1]{fontenc}

\DeclareRobustCommand{\VAN}[3]{#2}
\let\VANthebibliography\thebibliography
\def\thebibliography{\DeclareRobustCommand{\VAN}[3]{##3}\VANthebibliography}

\usepackage{graphicx}	
\usepackage{amsmath}	
\usepackage{caption}
\usepackage{gensymb}
\usepackage{adjustbox}
\usepackage{tablefootnote}
\usepackage{float}
\usepackage{ragged2e}
\usepackage{tabularx}

\newcommand{\hst}{\textit{HST}}
\newcommand{\astrosat}{\textit{AstroSat}}

\newcommand{\swift}{\textit{Swift}}
\newcommand{\nicer}{\textit{NICER}}

\newcommand{\source}{MAXI J1820+070}

\newcommand{\Msun}{\mathrm{M}_{\odot}}
\newcommand{\lum}{\mathrm{erg~s}^{-1}}

\newcommand{\mdot}{\mathrm{M_{\odot}~yr}^{-1}}

\newcommand{\kms}{\mathrm{km~s}^{-1}}

\newcommand{\ditto}[1][.4pt]{\textquotedbl}

\title[UV spectroscopy across an LMXB state transition]{Ultraviolet spectroscopy of the black hole X-ray binary MAXI J1820+070 across a state transition}
\author[M. Georganti et al.]{M. Georganti,$^{1}$\thanks{E-mail: Maria.Georganti@soton.ac.uk}
C. Knigge, $^{1}$ 
N. Castro Segura,$^{2,1}$
K. S. Long,$^{3,4}$ 
G. C. Dewangan,$^{5}$
S. Banerjee,$^{5}$
\newauthor R. I. Hynes,$^{6}$ 
P. Gandhi, $^{1}$ 
D. Altamirano, $^{1}$
J. Patterson, $^{7}$ and
D. R. Zurek $^{8}$
\\
$^{1}$ School of Physics and Astronomy, University of Southampton, Southampton, Hampshire, SO17 1BJ, UK \\
$^{2}$ Department of Physics, University of Warwick, Coventry CV4 7AL, UK \\
$^{3}$ Space Telescope Science Institute, 3700 San Martin Drive, Baltimore, MD 21218, USA \\
$^{4}$ Eureka Scientific, Inc.2452 Delmer Street, Suite 100, Oakland, CA 94602-3017, USA \\
$^{5}$ Inter-University Centre for Astronomy and Astrophysics (IUCAA), Ganeshkhind, Pune 411007, India \\
$^{6}$ Department of Physics and Astronomy, Louisiana State University, Baton Rouge, LA 70803, USA \\
$^{7}$ Department of Astronomy, Columbia University, 538 West $\&$ 120th Street, New York, NY 10027, USA \\
$^{8}$ Department of Astrophysics, American Museum of Natural History, CPW $\&$ 79th street, New York, NY 10024-5192, USA
}
\date{Accepted XXX. Received YYY; in original form ZZZ}

\pubyear{2025}

\begin{document}
\label{firstpage}
\pagerange{\pageref{firstpage}--\pageref{lastpage}}
\maketitle

\begin{abstract}
We present ultraviolet (UV) spectroscopic observations covering three distinct accretion states of the low-mass X-ray binary (LMXB) MAXI J1820+070: the luminous hard state, a hard-intermediate state and the soft state. Our observations were obtained during the 2018 eruption of MAXI J1820+070 with the {\em Hubble Space Telescope} (\hst) and \astrosat{} observatory. The extinction towards the source turns out to be low -- $\rm E_{B-V} = 0.2 \pm 0.05$ -- making it one of the best UV accretion laboratories among LMXBs. Remarkably, we observe only moderate differences between all three states, with all spectra displaying similar continuum shapes and emission lines. Moreover, the continua are not well-described by physically plausible irradiated disc models. All of this challenges the standard reprocessing picture for UV emission from erupting LMXBs. The UV emission lines are double-peaked, with high-ionization lines displaying higher peak-to-peak velocities. None of the lines display obvious outflow signatures, even though blue-shifted absorption features have been seen in optical and near-infrared lines during the hard state. The emission line ratios are consistent with normal abundances, suggesting that the donor mass at birth was low enough to avoid CNO processing ($\rm M_{2,i} \lesssim 1.0 - 1.5 {\mathrm M_{\odot}}$). Finally, we study the evolution of UV variability in our time-resolved \hst~observations (hard and hard-intermediate states). All UV power spectra can be modelled with a broken power-law, superposed on which we tentatively detect the  $\simeq 18$~s quasi-periodic oscillation (QPO) that has been seen in other spectral bands. 
\end{abstract}




\section{Introduction}

\source~(=ASASSN-18ey) is one of the closest, brightest and least-reddened Galactic black hole X-ray transients (BHXTs). It was discovered after its eruption on 2018 March \citep{tucker18,kawamuro18}. Since it is an excellent target for studying black hole (BH) accretion, unprecedented monitoring has been carried out across the entire electromagnetic spectrum, covering the main long outburst of the source and the subsequent reflaring episodes \citep[e.g.][]{baglio18,atri20, stiele2020, tetarenko21a, arabaci22, yoshitake22,banerjee24}. 

During their outbursts, BHXTs typically follow the same evolutionary track and cycle between two distinctive accretion states, the \textit{hard} and the \textit{soft} state. The outburst evolution is hysteretic: in the  \textit{hardness-intensity diagram} \citep[HID; e.g. see][for more information]{fender04}, where the spectral hardness of the accretion flow is plotted against the X-ray luminosity, erupting BHXTs trace a "q-shaped" path. The geometry of the accretion flow, the X-ray spectrum and the variability properties are all strongly \textit{state-dependent} (e.g. \citealt{esin97, homan01, zdziarski04, homan05, remillard06,done07,belloni10, gilfanov10, belloni16} and references therein).

At the beginning and end of an outburst, the system is in the hard state. In this state, the inner edge of the geometrically thin accretion disc is located far from the BH. The broadband spectral energy distribution (SED) is then dominated by a hard (power-law, $\rm \Gamma < 2$, high-energy cut-off around 100 keV) component, which is the result of low-energy photons being upscattered by relativistic electrons in a hot "corona". A weak thermal component associated with the accretion disc may also be apparent in lower energies \citep{grove98}. Timing observations tend to show strong X-ray variability $-$ of the order of 20-50 per cent \citep[e.g.][]{belloni05, munoz-darias11, belloni14, motta16} $-$ while the power density spectra (PDS) are characterized by red noise, low-frequency breaks, and often, quasi-periodic oscillations (QPOs). The latter are mostly seen in the low-frequency regime (0.01-30 Hz) and commonly associated with either geometrical effects such as the Lense-Thirring precession of the hot accretion flow in the vicinity of the BH \citep{stella98, stella99, schnittman06, ingram09, ingram19}, instabilities such as the accretion/jet instability \citep{tagger99, varniere02, varniere12, ferreira22} or coronae oscillations \citep{titarchuk04, cabanac10}. At radio and infrared wavelengths, collimated, relativistic jets are also commonly observed in this state \citep[e.g.][]{markoff01, corbel03, fender04, fender09, fender14}.

As the source brightens and the edge of the inner disc moves toward the innermost stable circular orbit (ISCO) of the BH, the system proceeds to the \textit{soft} state. Notable changes occur in the X-ray spectrum. The hard component becomes softer ($\rm \Gamma > 2$) but characteristically extends in a coronal tail exceeding 500 keV associated with the faint Comptonization region \citep[e.g.][]{belloni16} while the SED is dominated by the soft thermal component associated with the geometrically thin, optically thick accretion disc \citep{shakura73}. 
Variability during the soft state is limited to amplitudes of a few per cent \citep[commonly between 1-5 per cent; e.g.][]{casella04, belloni05, munoz-darias11, belloni14, motta16}, while the PDS are  described by a weaker, broken, power-law component. QPOs are also sometimes present, but they are broader and weaker than the ones found in the hard state. They are plausibly associated with the accretion/jet instability \citep{varniere02, varniere12} or the launch of the relativistic jet away from the system \citep{fender09}. Radio emission is quenched in the soft state \citep[e.g.][]{fender99, corbel01, fender04, miller08,fender09, russell11}, but blue-shifted X-ray absorption lines associated with highly ionized disc winds often emerge \citep[e.g.][]{miller06b, miller08, neilsen09, ponti14,  ponti16, diaz-trigo16, higginbottom18}, preferentially seen in high-inclination \citep{ponti12} and sub-Eddington luminosity \citep{higginbottom19, higginbottom20} systems. The transitions between hard and soft states proceed via \textit{intermediate} states, which can arise at different luminosities due to the hysteretic nature of the eruptions. However, these transitional states are more difficult to study and can present complex signatures during their brief window of occurrence.

\begin{figure*}
\includegraphics[width=\textwidth]{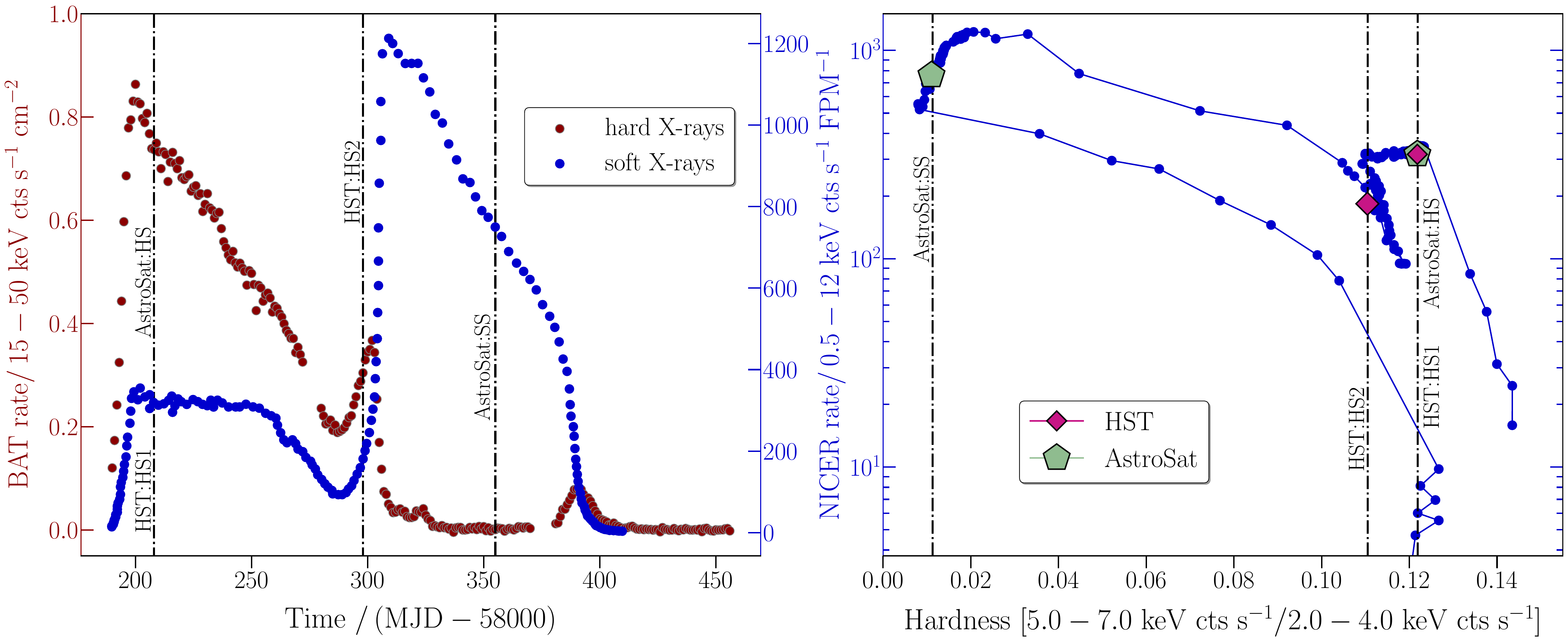}
\caption{{\em Left panel}: \swift/BAT (hard X-rays) and \nicer~(soft X-rays) light curves of the outburst of \source~in 2018. {\em Right panel}: \nicer~HID of the source (blue), averaged by day. In both cases, we have superimposed the timing of our UV observations for comparison. The times of our \hst~(pink) and \astrosat~(light green) observations are noted by different colour, signifying the respective accretion state of the system. The two panels share the same label as they are referred to \nicer~observations. The transition between the two accretion states is evident in both panels.}
\label{fig:X-ray_lc_HID}
\end{figure*}

A significant challenge for the study of BHXTs is that interstellar extinction usually prevents us from exploring the ultraviolet (UV) response to changes in X-ray luminosity and state transitions. This is unfortunate, as the UV band is a critical link between the inner accretion flow (the corona and the hot, viscously-dominated disc) and the outer one (the cool, irradiation-dominated disc). To date, only six BHXTs have been observed in the far-UV part of the spectrum near outburst peak in the soft state \citep{hynes05}. Increasing the size of this sample is important for understanding how the disc evolves and responds to the changing X-ray irradiation during eruptions, what drives the presence of disc wind signatures in specific wavelengths during different states \citep[e.g.][]{munoz-darias16, munoz-darias18, munoz-darias19, sanchez-sierras20, castro-segura22, fijma23}, and to shed light on the periodic and aperiodic variability produced at intermediate disc radii. \source~provides us with a rare opportunity to make progress in this area. Here, we present the first multi-epoch UV characterization of a BHXT across both outburst states, which allows us to shed light on some of these issues. 

The outbursts of BHXTs are typically interpreted in the context of the \textit{disc instability model} (DIM; e.g. \citealt{lasota01, hameury20} for a recent review), according to which the accretion discs in these systems are subject to a thermal-viscous instability related to the ionization of hydrogen. To what extent X-ray irradiation plays a significant role in controlling the behaviour of the discs in BHXTs is still a matter of debate \citep[e.g.][]{vanparadijs94, vanparadijs96,king98, dubus99, dubus01, tetarenko18b}. Strongly irradiated accretion discs are needed, though, to explain the longer and brighter eruptions of these systems and their light curve phenomenology. The outer disc is heated by radiation emitted in the inner regions, which kepdf it ionized and prevents an early return to quiescence. It also means that the outer disc produces most of the UV/optical, and sometimes infrared, flux. This picture can be tested by modelling the UV SED with irradiated disc models.

The spectral lines seen in the UV also provide important diagnostics. For example, the \ion{He}{ii} $\lambda 1640$\AA~recombination line is an effective bolometer for the (usually unobservable) extreme UV (EUV) band above 54 eV. Moreover, UV line \textit{ratios} can be used to shed light on the evolutionary history of the binary, since they strongly depend on the degree of CNO processing the accreting material has been subjected to \citep[e.g.][for both the cases of cataclysmic variables, CVs, and LMXBs]{mauche97, haswell02, gansicke03, froning11, froning14, castro-segura24}. Perhaps most importantly, however, the strong UV resonance lines (\ion{N}{v} $\lambda 1240$\AA, \ion{Si}{iv} $\lambda 1400$\AA, \ion{C}{iv} $\lambda 1550$\AA) are highly sensitive outflow tracers.

The observable wind signatures presented by spectral lines take the form of blue-shifted absorption or P-Cygni profiles. In LMXBs, such signatures have been found in both the corona- and disc-dominated states. However, clear \textit{X-ray} wind signatures have so far only been seen in the soft state \citep[e.g.][and references therein]{miller06a, ponti12, ponti14, diaz-trigo16}, even though hard state outflows have been identified both in optical/near-infrared \citep[e.g.][]{munoz-darias16, munoz-darias18, munoz-darias19, jimenez-ibarra19} and UV spectra \citep{castro-segura22, fijma23}. This raises several important questions \citep[e.g.][]{munoz-darias19, sanchez-sierras20, munoz-darias22}. Is the same wind present throughout the eruption, perhaps in different ionization states? And is this wind multi-phase (e.g. clumpy) and/or spatially stratified? Or are we seeing completely distinct types of outflows in different bands (e.g. cool, magnetically driven winds in the UV, optical and near-infrared bands vs hot, thermally driven winds in X-ray bands)? Resolving these issues will require panchromatic spectroscopic observations across different spectral states. In this context, the present study provides the first data set allowing a search for UV wind signatures in different stages of a single outburst. 

Finally, time variability in the form of stochastic flickering and QPOs is ubiquitous among X-ray binaries (and, in fact, among all accreting systems). The standard interpretation of this variability is that each annulus of the disc is susceptible to fluctuations on its own viscous timescale. These fluctuations then propagate inwards, so that the correlated variability we see in different wavelengths is the cumulative product of variations across a wide range of disc radii. Irradiation and reprocessing can introduce additional correlations between different wavebands on timescales corresponding to the light travel time from the center of the disc to the relevant emitting region. \source~is already known to display correlated variability between X-rays and optical at the peak of its outburst \citep[e.g.][]{stiele2020, paice21, thomas22}. Our UV data allows us to test whether / to what extent the same behaviour is present in this intermediate waveband. 

\source~reached peak optical brightness around March 28, but remained in the hard state until (early) July. The system then started its transition to the soft state, characterized by a rapid softening of the X-ray spectrum \citep{homan18a, homan18b, homan20} and a decrease of radio and infrared flux, thereby indicating quenching of the compact jet \citep{tetarenko18a, casella18}. In this work, we focus on three distinct snapshots of \source~obtained during this outburst with the \textit{Hubble Space Telescope} (\hst) and \astrosat~observatories. The timing of our observations is shown in Figure \ref{fig:X-ray_lc_HID}, where we relate them to the overall X-ray behaviour of the system. For this purpose we have combined data from both the \textit{Neil Gehrels Swift Observatory} (\swift/BAT; \citealt{gehrels04, krimm13}) and  \textit{Neutron star Interior Composition Explorer} (\nicer; \citealt{gendreau16}) to construct both the overall X-ray light curve and the HID of the system.

\section{Observations and data reduction}

\begin{table*}
\caption{Log of observations discussed in this work.  The last column represents the notation that we are going to use throughout this paper.}\label{tab:log}
\begin{threeparttable}
\begin{tabular*}{\textwidth}{@{\extracolsep{\fill}} ccccccc}
\hline 
\hline
Observatory & ObsID & Setup & Onset (MJD) & Exposure (ks) & Accretion state & Notation \\
\hline 
\hst & ods801010 & STIS/E140M/1425\AA & 58208.52 & 2.151 & & \\
& ods801020 & \ditto & 58208.57 & 2.730 & & \\
& ods801030 & \ditto & 58208.64 & 2.730 & Hard & HST:HS1 \\
& ods801040 & STIS/E230M/1978\AA & 58208.71 & 2.730 & & \\
& ods801050 & STIS/E230M/2707\AA & 58208.77 & 2.730 & & \\
\hline
\hline
\hst & ods802010 & STIS/G140L/1425\AA & 58298.50 & 2.105 & &   \\
& ods802020 & \ditto & 58298.56 & 2.730 & & \\
& ods802030 & \ditto & 58298.63 & 2.730 & Hard & HST:HS2 \\
& ods802040 & STIS/G230L/2376\AA & 58298.70 & 2.730 & & \\
& ods802050 & \ditto & 58298.76 & 2.670 & & \\
\hline
\hline
\astrosat & T02$\_$038T01$\_$900001994 & UVIT/FUV-G1 & 58207.50 & 11.39 & Hard  & AstroSat:HS \\
\hline
\hline
\astrosat & T02$\_$066T01$\_$900002324 & UVIT/FUV-G1 & 58355.47 & 2.845 & Soft & AstroSat:SS \\ 
\hline  
\hline
\end{tabular*}
\label{tab:observations}
\end{threeparttable}
\end{table*}

\subsection{Hubble Space Telescope}
\source~was observed on 2018 March 31st and June 29th (Proposal ID: 15454, PI: Knigge) by the  \textit{Space Telescope Imaging Spectrograph} \citep[STIS;][]{kimble98, woodgate98} onboard \hst. Each epoch consisted of a five-orbit \hst~visit, with three orbits dedicated to the far-UV and the remaining two to the near-UV part of the spectrum. Both visits cover the \textit{hard} state,  albeit at quite different stages of the outburst evolution. The first visit took place right after the outburst peak, while the second took place three months later, just before the start of the hard-to-soft state transition. 

The MAMA detectors were employed to obtain the observations in \textsc{time-tag} mode in order to provide information on the fast (dynamical timescale) variability of the system. The initial observation, denoted as HST:HS1, captured the system in a luminous hard state on MJD=58208.5, just few days after the outburst peak. The $0\arcsec.2\times0\arcsec.2$ detector slit and the  \textit{echelle} E140M/1425\AA, E230M/1978\AA, 2707\AA~gratings were utilized, providing us with a spectral resolution of $\mathrm R=\lambda$/$\Delta \lambda$ = 45800 and 30000 in the far- and near-UV regions, respectively. The subsequent observation, designated as HST:HS2, took place on MJD=58298.5, just days before the state transition. In this epoch, we used the $52\arcsec\times0\arcsec.2$ slit coupled with the \textit{first-order} G140L/1425\AA~and G230L/2376\AA~gratings. Here, the spectral resolution is not constant but wavelength-dependent on both sides of the spectrum. In particular, it ranges between 960-1440 in the far-UV and between 500-1010 in the near-UV wavelengths, respectively.  A summary of the instrumental setups and their characteristics can be found in Table \ref{tab:log}. 

All the data were reduced using the \textsc{calstis} pipeline to extract one-dimensional spectra for each visit. For our first epoch, blaze correction of the individual orders of the echelle spectra and trimming of their overlapping regions were applied so that we could combine them into a single spectrum. On the other hand, the construction of the HS2 first-order spectrum suffered from two challenges. First, there was a flux inconsistency between the far- and near-UV exposures $-$ in particular, the near-UV (G230L) ods802050 exposure appeared to be 27$\%$ brighter than the corresponding far-UV (G140L) one in the overlap regions between the two. We suspect that this mismatch is an instrumental artefact, possibly associated with the slight offset in the wavelength calibration discussed below (which can affect the flux level via the wavelength-dependent sensitivity curve). While we cannot completely rule out that this offset is real (i.e. produced by intrinsic variability), we adjusted the near-UV time-weighted average to match its far-UV counterpart. This adjustment involved determining a scaling factor based on the median value of continuum windows on both sides of the spectrum. The purpose of this choice was to facilitate our subsequent analysis by creating a single HS2 spectrum. Second, we observed a slight mismatch between the far- and near-UV line centers. As the setup in this epoch is constrained by the lack of spectral resolution and low signal-to-noise ratio, we used the echelle spectrum and the location of several interstellar lines as our guide to account for these wavelength shifts. In the end, we shifted the far-UV wavelengths by 1\AA~and the near-UV wavelengths by 2.5\AA~to correct for these offsets. 

\subsection{AstroSat}
The \astrosat~satellite observed our source on March 31st (hard state) and August 25th (soft state), utilizing its \textit{Ultraviolet Imaging Telescope} \citep[UVIT;][]{tandon2017orbit,tandon2020additional}. Here, these epochs are denoted as AstroSat:HS and AstroSat:SS, respectively.  

The UVIT instrument features three channels sensitive to three different bands: far-UV (1200-1800\AA), near-UV (2000-3000\AA) and visible (3200-5500\AA). While the far- and near-UV are used for scientific observations, the visible channel primarily aids in pointing drift correction. Both the UV channels are equipped with broadband filters for imaging, providing a point spread function (PSF) of $1.0\arcsec$-$1.5\arcsec$, and include slit-less gratings for low-resolution spectroscopy. The far-UV channel contains two orthogonally arranged slit-less gratings (FUV-G1 and FUV-G2) to minimize contamination along the dispersion direction from nearby sources. Both channels operate in the photon counting mode. Further details on grating performance and calibration can be found in \cite{dewangan2021}.

Level-1 data on \source~were sourced from the AstroSat archive and processed with the \textsc{ccdlab} pipeline \citep{postma2017}. Drift-corrected, dispersed images were generated orbit-wise and then aligned to produce a single image per observation. Spectral extraction was performed using the \textsc{uvittools}  package\footnote{Details about the package's requirements and documentation can be found at {\small \url{https://github.com/gulabd/UVITTools.jl}.}}, following \citet{dewangan2021} and \citet{kumar2023}. The source's zeroth-order image position in grating images was located, and one-dimensional count spectra for the far-UV gratings in the -2 order were extracted using a 50-pixel cross-dispersion width (see \cite{banerjee24} for more details). Background count spectra were also extracted similarly from source-free regions and used to correct the source spectra. The grating responses were updated to match the simultaneous hard state \hst~spectrum of the source and these files are produced as explained in \cite{dewangan2021}.

\section{Spectroscopic analysis}  \label{sec:Spectroscopic_analysis}

\subsection{Reddening and HI column density}
\label{sec:reddening}

Most BH binaries are located at kiloparsec (kpc) distances in the Galactic plane. As a result, they tend to suffer from strong extinction and reddening, making them difficult to observe especially at UV wavelengths. In fact, only 8 out of 68 BH binary systems are characterized by reddening values $\rm E_{\rm B-V} \leq 0.3$ \citep{corral-santana16}. In addition to \source~(whose extinction properties we discuss below), these low-reddening systems are GRS 1009-45 \citep{DellaValle97}, XTE J1118+480 \citep{Garcia00}, Swift J1357.2-0933, XTE J1817-330 \citep{Schlafly11}, MAXI J1305-704 \citep{matasanchez21} and MAXI J0637-430 \citep{tetarenko21b}. Thanks to their relatively low and well-constrained extinction values, the intrinsic luminosities of these systems can be inferred with some confidence.

\begin{figure}
\centering
\includegraphics[width=0.5\textwidth]{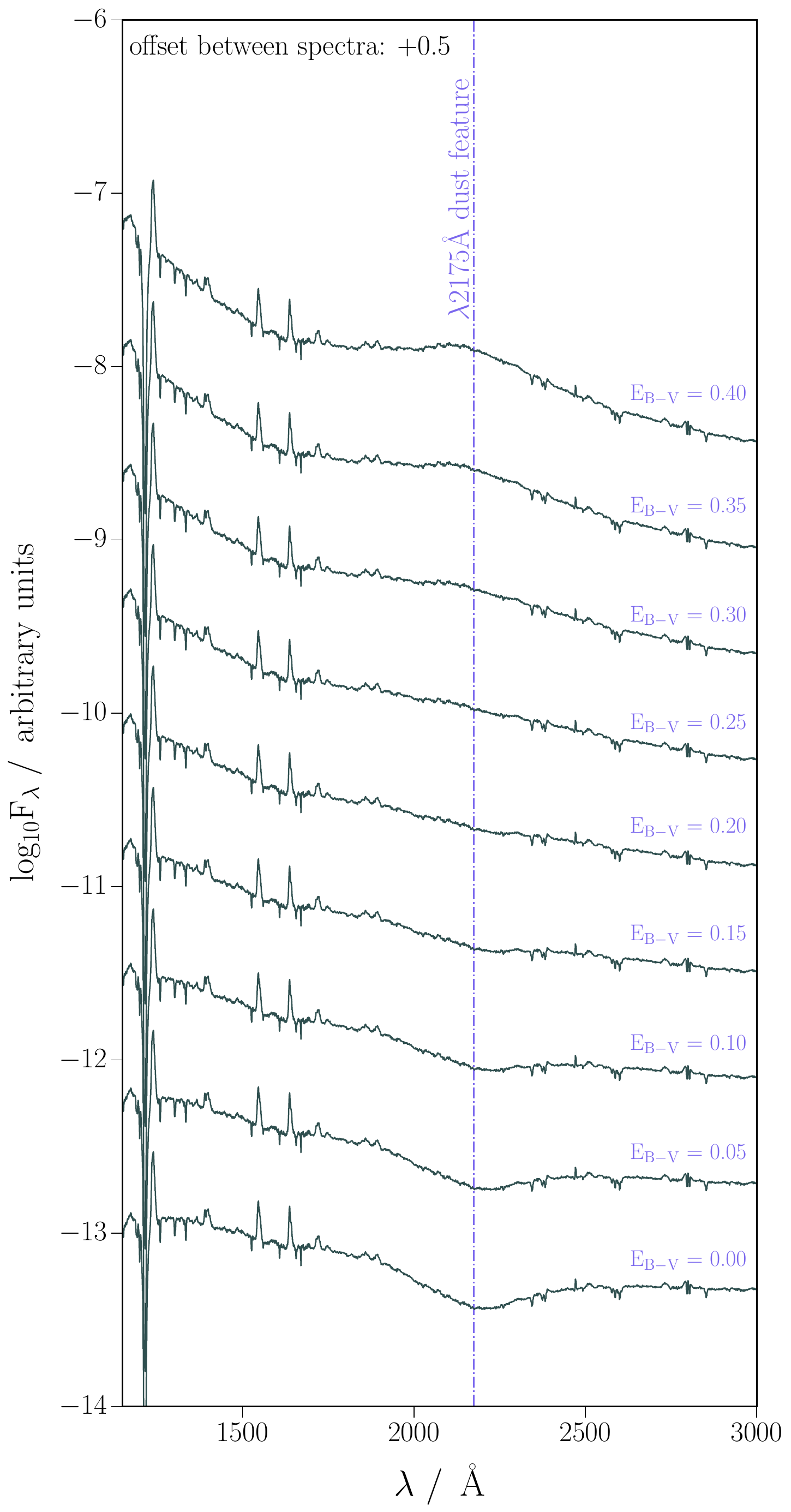}
\caption{Modelling of the HST:HS2 spectrum to estimate the line-of-sight reddening to \source. We employ the strength of the $\lambda$2175\AA~interstellar feature and Fitzpatrick's extinction law \citep{fitzpatrick99} to model the HST:HS2 spectrum and find the reddening value, which makes the near-UV bump to disappear. From this analysis, we derive an interstellar reddening value of $\rm E_{B-V}=0.2$, which is the one adopted in this paper.}
\label{fig:MAXI_J1820_F99_extinction}
\end{figure}

In order to determine the line-of-sight extinction and reddening towards \source, we use two different methods. First, we fit a series of parameterized extinction curves to the UV SED. These extinction curves are characterized by the parameters $\rm A_{v}$ and $\rm R_{v}$, where $\rm E_{\rm B-V}$=$\rm A_{v}/R_{v}$. Here, we set $\rm R_{v}$=3.1, the standard value for Galactic sources. Using Fitzpatrick's reddening-law \citep{fitzpatrick99} to match the $\lambda$2175\AA~interstellar dip of our "featureless"/continuum HST:HS2 spectrum, we estimate an extinction value of $\rm E_{\rm B-V}= 0.20 \pm 0.05$. The quoted uncertainty here is a qualitative, but conservative estimate of the systematic uncertainty associated with this procedure. Figure \ref{fig:MAXI_J1820_F99_extinction} illustrates how the strength of the near-UV bump changes for different extinction values.

In addition, we exploit the well-established correlation between reddening and neutral atomic hydrogen column density ($\rm N_H$) at any line-of-sight \citep[e.g.][]{bohlin78, liszt14}. In our case, we determine the HI column density by modelling the damped Lyman-$\alpha$ (Ly$\alpha$) absorption profile, evident in our HST:HS1 spectrum. We follow \citet{bohlin75} and model this  profile by an $\rm C e^{-\tau}$ factor, where C is the continuum level, and the optical depth, $\tau$, is given by
\begin{equation}
\tau (\lambda) = 4.26 \times 10^{-20} \rm N_H/(6.04 \times 10^{-10} + (\lambda - \lambda_0)^2).
\end{equation}

\begin{figure}
\includegraphics[width=0.49\textwidth]{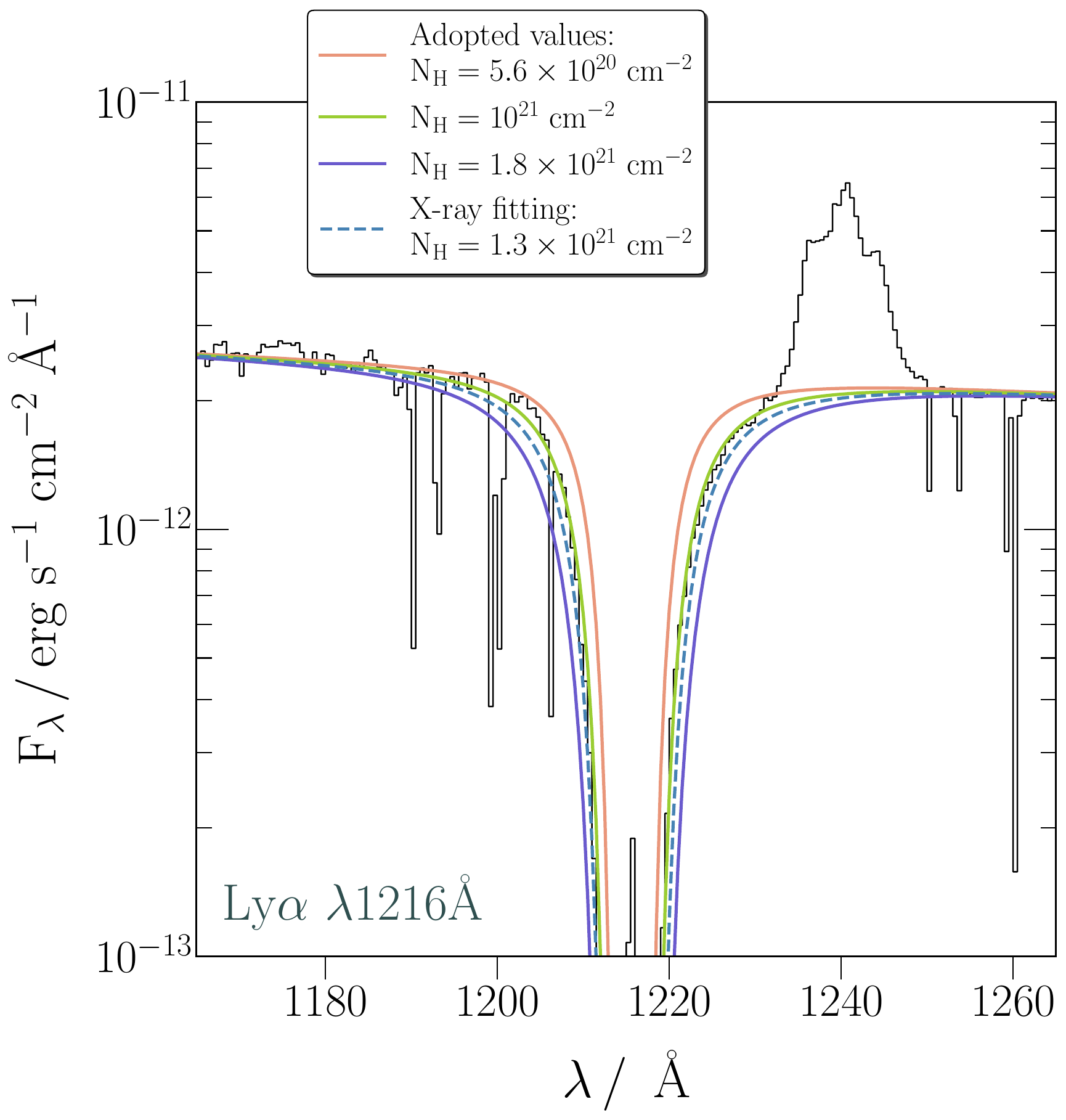}
\caption{Modelling of the damped profile of the interstellar Ly$\alpha$ line in our HST:HS1 spectrum. The different fits are represented by different solid coloured lines, whose corresponding column density is specified in the legend. The light blue dashed line signifies the column density, derived from X-ray spectral fitting, as described by \citet{koljonen23}. The optimal model, represented by a light green colour, corresponds to a column density of $\rm N_H=10^{21} cm^{-2}$.}
\label{fig:Lya_modelling}
\end{figure}

Here, $\mathrm \lambda_0 = 1215.67$\AA, the rest wavelength of the Ly$\alpha$ line. We then estimate $\rm N_H$ by trial and error within a range of values centered on the total line-of-sight column density. Examples of these fits are shown in Figure \ref{fig:Lya_modelling}, where the best-fit model corresponds to a column density value of $\rm N_H = 10^{21}\ cm^{-2}$. The corresponding reddening value then is determined via two standard $\rm N_H/E_{B-V}$ relations from the literature: $\rm N_H/E_{B-V} = 5.8 \times 10^{21}\ cm^{-2}\ mag^{-1}$ \citep{bohlin78} and $\rm N_H/E_{B-V} = 8.3 \times 10^{21}\ cm^{-2}\ mag^{-1}$ for $|b|<30 \degree $ \citep{liszt14}. The resulting estimates for $\rm E_{\rm B-V}$ towards \source~are 0.17 and 0.12, respectively. These values are slightly lower, but remain consistent with the one inferred from the UV SED. 

Similar reddening results are obtained when employing slightly different $\rm N_H$ values derived from other observational studies. For example, high-resolution 21-cm radio observations \citep{h14pi_collaboration16} and fitting of the absorbed X-ray spectrum \citep{koljonen23} yield a column density of $\rm N_H=1.3\times10^{21} cm^{-2}$. On the other hand, modelling of diffuse interstellar bands (such as the $\lambda5780$\AA~line in {\em VLT}/X-Shooter spectra) results in a column estimate of $\rm N_H=(1.4\pm 0.4)\times 10^{21} cm^{-2}$ \citep[also in][]{koljonen23}. These estimates are  consistent with predictions based on the Galactic dust distribution (see \citealt{koljonen23} and references therein). Throughout this work, we adopt $\rm E_{B-V}=0.2$ as the reddening value.

\subsection{Overview: spectroscopic evolution through the outburst} \label{sec:spectroscopic_overview}

\begin{figure*}
\includegraphics[width=\textwidth]{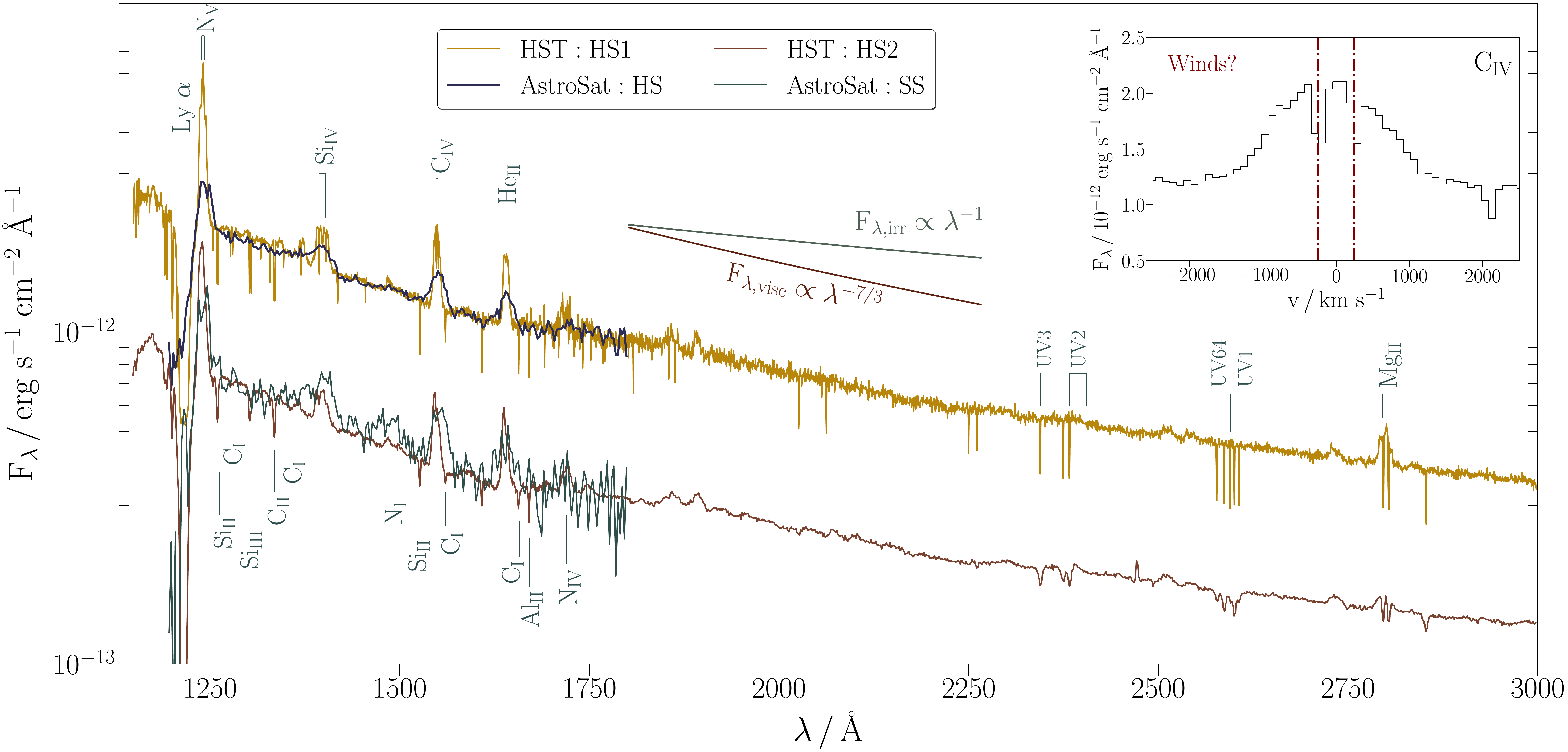}
\caption{The dereddened time-averaged UV spectra of \source~as the system evolves across three distinct stages of its outburst, accompanied by line identifications of the most prominent species. Our work covers three key points of the source's outburst: a) a luminous hard state, combining (quasi-)simultaneous \hst/\astrosat~observations b) \hst~observations of a lower luminosity hard state before the state transition and c) \astrosat~observations of the soft state. The \hst~spectra cover the far- and near-UV regions (1150-3000\AA) whereas the \astrosat~observations cover only the 1200-1800\AA~regime. As a reference, the slope indices of both a standard Shakura-Sunyaev accretion disc ($\rm F_{\lambda, visc} \propto \lambda^{-7/3}$) and an irradiated disc ($\rm F_{\lambda, irr} \propto\lambda^{-1}$) are illustrated. For clarity, we cut the interstellar lines by 70$\%$ of their flux, compared to the original lines. The inset zooms in on the \ion{C}{iv} $\lambda1550$\AA~profile, emphasizing the absence of an outflow.}
\label{fig:MAXI_J1820_spectral_evolution}
\end{figure*}

Our first objective is to shed light on the UV spectral evolution of \source~as it passes from the hard to the soft state during its 2018 outburst. The luminous hard state near the outburst peak was covered by (quasi-)simultaneous \hst~and \astrosat~observations. The less luminous hard state just before the state transition was caught by our second \hst~visit, while the soft state was captured only by low-resolution \astrosat~spectroscopy. The evolution of the source's spectrum among these three stages is shown in Figure \ref{fig:MAXI_J1820_spectral_evolution}.

Overall, the \hst~spectra are characterized by blue continua and broad emission lines. Despite the significant flux decrease, and the system's transition from the hard to the soft state, immediately after the second \hst~epoch, no major characteristic differences are observed between these two observations. The continuum shape remains consistent across our spectra and is reasonably well-approximated by a power-law with an index close to the characteristic value for a viscously-dominated accretion disc ($\rm F_{\lambda,visc} \propto \lambda^{-\beta}$, $\mathrm \beta=7/3 $). This is surprising, since the disc regions producing the UV continuum may be expected to be heated by irradiation, rather than viscous dissipation. At wavelengths sufficiently far away from the Wien and Rayleigh-Jeans tails, an irradiation-dominated disc would tend to produce a power-law continuum with $\beta=1$, significantly shallower than observed. We will return to this issue below in Section \ref{sec: Irradiated_disc_modelling}.

Both epochs display the same set of strong, double-peaked emission lines, representing both low- and high-ionization atomic transitions. At shorter wavelengths, we see the usual resonance UV lines such as \ion{N}{v} $\lambda1240$\AA, \ion{Si}{iv} $\lambda1400$\AA, \ion{C}{iv} $\lambda1550$\AA~or \ion{He}{ii} $\lambda1640$\AA~in emission. The damped Ly$\alpha$ absorption line produced by absorption in the interstellar medium (ISM) contaminates the far-UV end of the spectrum. At the very opposite end, the prominent \ion{Mg}{ii} $\lambda2800$\AA~line in emission is apparent. 

\begin{figure}
\centering
\includegraphics[width=0.49\textwidth]{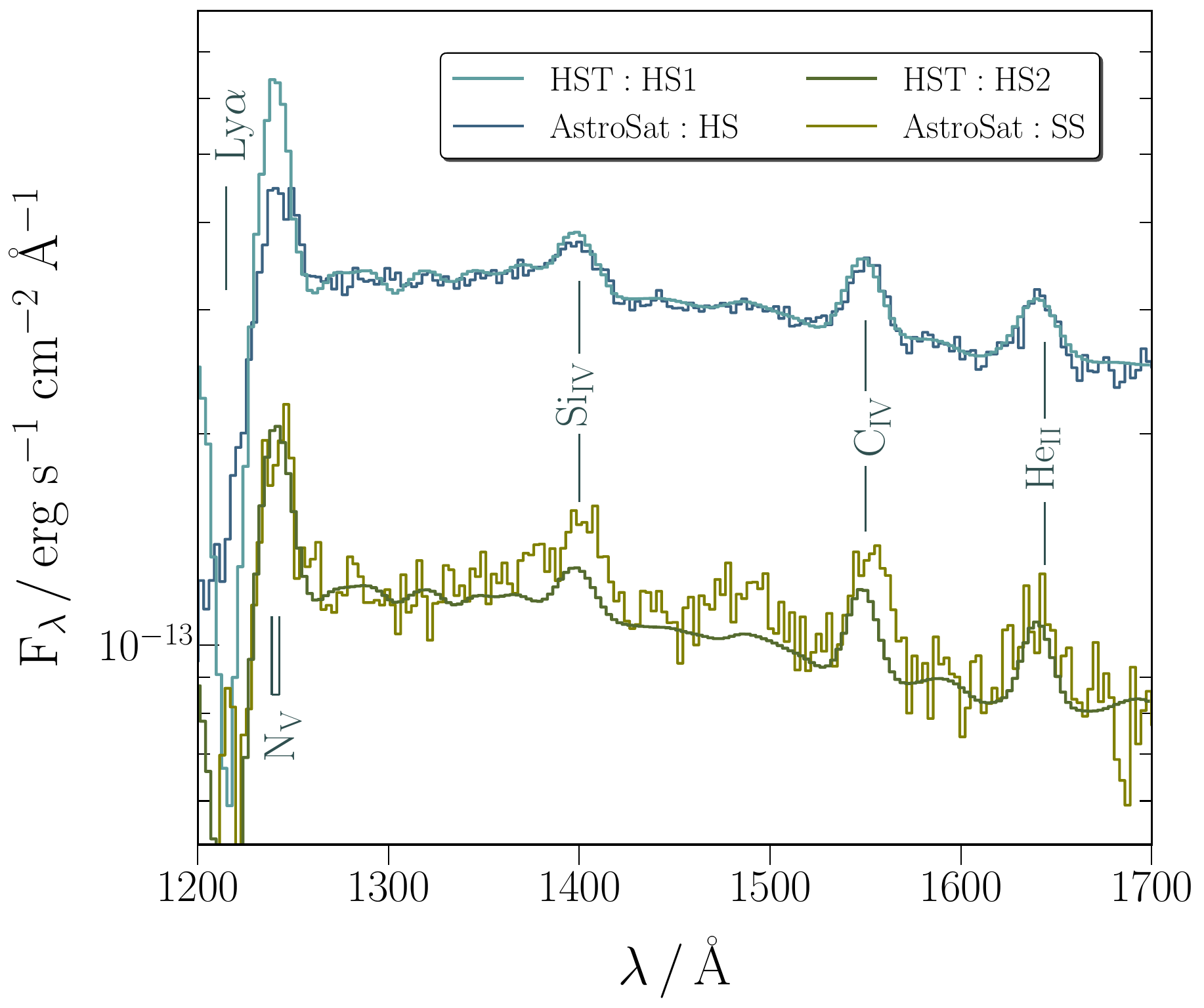}
\caption{Far-UV spectral evolution of \source~as it is captured by \hst~and \astrosat. We adjusted \hst's resolution of our hard state epochs to match \astrosat's intrinsic resolution. There is apparent agreement among all the observations, taken during the hard state, whereas we observe stronger features when the system is in the soft state. We note that we have limited information regarding the line profiles in the soft state, where only three lines are visible.}
\label{fig:MAXI1820_downgraded_spectra}
\end{figure}

The \astrosat~spectra also capture the far-UV wavelengths, but $-$ due to the lower resolution and signal-to-noise of these data sets $-$ only the strongest lines are detected. As \astrosat~constitutes our sole insight into the system's soft state, we exploit our simultaneous hard state observations to calibrate and facilitate comparisons between the \hst~and \astrosat~spectra. For these comparisons, we degrade the \hst~spectra to the same resolution as the \astrosat~observations using a simple Gaussian filter. The result is illustrated in Figure \ref{fig:MAXI1820_downgraded_spectra}. A good agreement is achieved between the (quasi-)simultaneous \hst/\astrosat~hard state observations, whereas there is a hint of line strengthening in the soft state, especially for the \ion{Si}{iv} and \ion{C}{iv} profiles.

\subsection{Emission line shapes and fluxes} \label{sec:line_fluxes}

All of the observed lines appeared to be double-peaked or consistent with being double-peaked (within the limit of the diverse spectral resolutions of our instrumental setups). Assuming that the observed emission lines are produced at the atmosphere of the accretion disc, their shapes can help us specify the physical conditions in the line-forming region(s) and define their evolution during the time between the observations. 

Broadened, double-peaked emission profiles are characteristic of geometrically thin Keplerian accretion discs, where the two peaks correspond to regions of the disc rotating towards (blue-shifted component) or away from the observer (red-shifted component), respectively. In this context, we carry out a qualitative analysis of the most prominent observed resonance lines from our dereddened continuum-subtracted spectra. Specifically, we measure properties such as their amplitude ratios and the corresponding velocities of each of the two components. Then, the fitted emission profile is numerically integrated to extract key parameters (like their line fluxes, $\rm F_{\lambda}$, and their equivalent widths, EWs). 

In our fits, all relevant \textit{atomic} transitions are assumed to produce two kinematically distinct (blue and red) Gaussian emission lines. The two components are constrained to share the same FWHM, but they are allowed to have different normalizations and shifts relative to the relevant rest wavelengths. This is not unexpected, as transient components such as hot spots or interactions at the stream-impact bulge can disrupt the disc's model symmetry. The \ion{He}{ii} line is adequately described by two such components associated with a single atomic transition. However, most of the lines we consider (specifically \ion{N}{v}, \ion{Si}{iv}, \ion{C}{iv} and \ion{Mg}{ii}) are resonance doublets, i.e. they consist of \textit{two} well-separated atomic transitions. We therefore have to explicitly account for both doublet components in our fits to these lines. Each of these lines is thus described by a total of four Gaussian emission profiles: an atomic doublet representing the blue kinematic component and an atomic doublet representing the red kinematic component. The wavelength separations between the atomic doublet components are always kept fixed at their known values. The intensity ratio between the doublet components is expected to range between 2:1 (in favour of the blue doublet component) and 1:1, depending on whether the line is optically thin or optically thick. However, we do not enforce this. 

\begin{figure*}
\includegraphics[width=\textwidth]{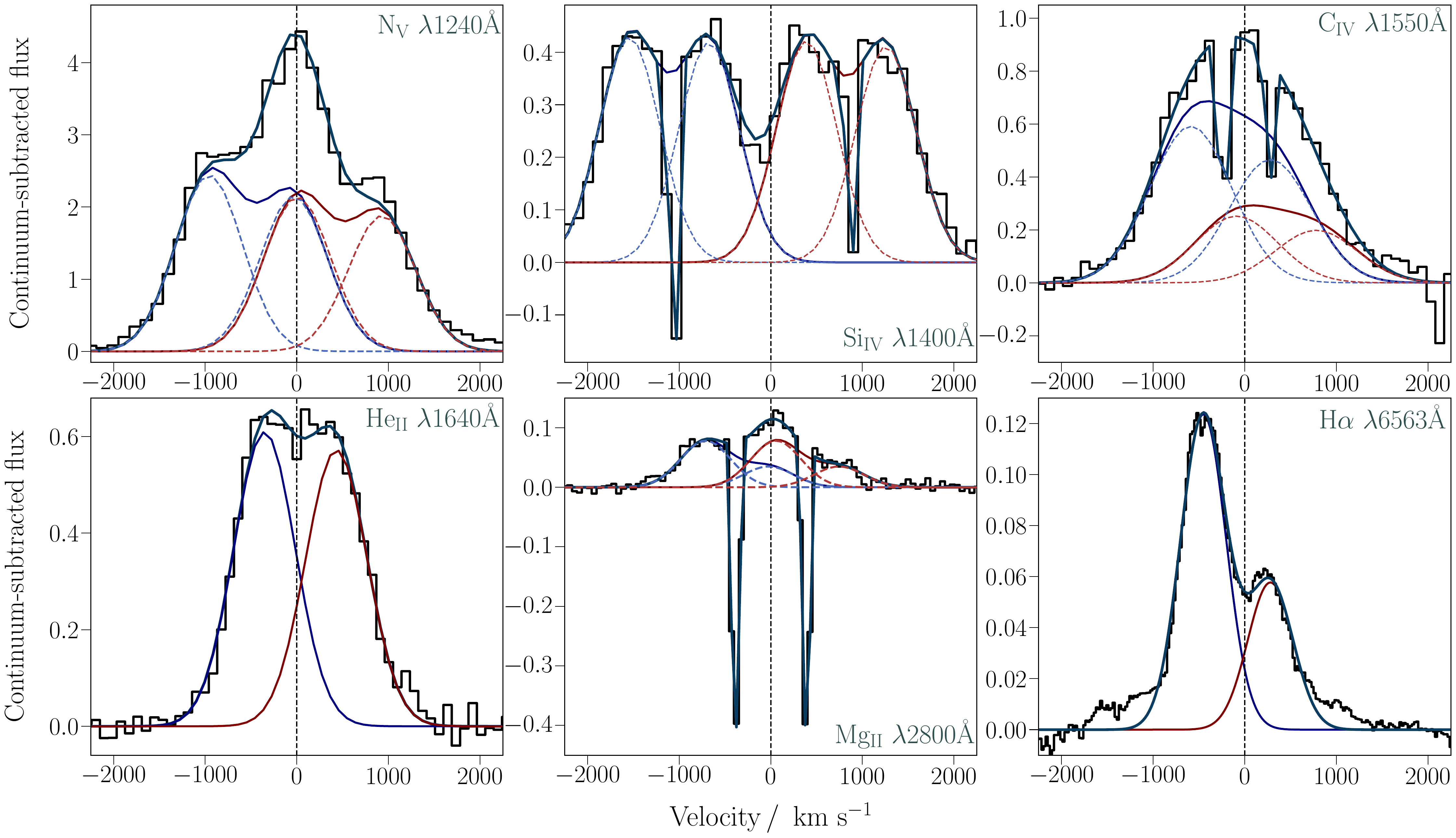}
\caption{Continuum-subtracted line profiles of the most prominent UV resonance lines, observed during our HST:HS1 epoch. Each profile is fitted by a number of Gaussian components to account for each of their individual shapes (dashed lines). Our modelling is based on the fact that each atomic transition produces a distinguised blue and a red emission component. Here, we follow the blue and red colour distinction to demonstrate the relevant created profiles. The last panel shows the \textit{VLT}/X-shooter $\rm H\alpha$ Balmer line, taken few days before the luminous hard state observation considered here, for comparison. Its profile demonstrates a hint of a shallow absorption blue wing, suggesting the presence of an outflow. Details of our Gaussian modelling are provided in Table \ref{tab: line_profiles}.}
\label{fig:MAXI_J1820_line_fits}
\end{figure*}

Our high-resolution echelle spectra obtained in epoch HST:HS1 (the luminous hard state) exhibit clear evidence for narrow absorption features close to the rest wavelengths of the \ion{Si}{iv}, \ion{C}{iv} and \ion{Mg}{ii} doublets. These absorption features might be intrinsic to the source, but are more likely associated with the ionized phase of the ISM along the line-of-sight. We therefore include narrow (unresolved) doublet absorption features in our models for these lines. The absorption profiles are modelled as $\rm e^{-\tau}$, where, here, $\tau$ corresponds to the superposition of the two absorption profiles. The profiles also share the same FWHM and their intensity ratio is specified to range between 2:1 (optically thin) and 1:1 (optically thick) cases. In particular, the only line considered to be formed in an optically thick region ($\tau >>1$) of the accretion disc is the \ion{Mg}{ii} line, as can be seen by its profile structure in Figure \ref{fig:MAXI_J1820_line_fits}. The same figure also shows the rest of the continuum-subtracted line profiles during our first \hst~epoch along with the $\rm H\alpha$ Balmer profile, taken by \textit{VLT}/X-shooter, days before our considered HST:HS1 date. The $\rm H\alpha$ profile shows shallow absorption at the blue wing, presumably of a moderate outflow flying away from the system. Its detailed analysis, though, and the reason of its appearance are out of the scope of the current paper. 

In the second \hst~epoch, described by lower resolution first-order spectra, the absorption profiles are not as strikingly  evident as in HST:HS1 and therefore, in our modelling, we fix their properties (amplitude, location and FWHM) to their already-determined values. It is noted that, for the \ion{Mg}{ii} line, in order to facilitate the modelling process, we also fix the intensity ratio and the (common) FWHM of the four-component Gaussian emission profile to the ones of the first \hst~epoch. In all cases, the model constructed of each of the lines is smoothed to the instrumental resolution of the employed setting and then, it is rebinned to 0.5\AA~(preferred data's binning). 

Finally, we implement the same method to extract line properties in \astrosat's dereddened spectra. However, in this case, we focus only on the strongest lines (\ion{Si}{iv} $\lambda1400$\AA, \ion{C}{iv} $\lambda1550$\AA, \ion{He}{ii} $\lambda1640$\AA) observed in our spectra. Due to the \astrosat's lower resolution, the line profiles of these species are satisfactorily fitted by two Gaussian components, except for the case of the \ion{He}{ii} line, which is well-modelled as just a single Gaussian. We avoid measuring the properties of the \ion{N}{v} profile, as it lies too close to the blue edge of the detector to yield reliable measurements. 

The line flux and EW uncertainties associated with our fits are estimated through Monte Carlo simulations. First, we estimate the standard deviation (rms), $\rm \sigma_{res}$, between the data and the fitted model. Then, for each line, we generate mock datasets, where the model flux at each wavelength is perturbed by a Gaussian distribution centered on the actual flux with standard deviation, $\rm \sigma_{res}$. The line flux and EW for each of these mock sets are then calculated by fitting the new line profile and numerically integrating across it. This iterative process is repeated 1000 times, allowing us to estimate the errors at 1$\rm \sigma$ confidence level. 

The results of our fitting process are presented in Table \ref{tab: line_profiles}. The peak-to-peak velocities, $\rm v_{pp}$, shown also in Figure \ref{fig:MAXI1820_p2p_veloc_radii_potential}, are consistent with the standard picture of an accretion disc where higher ionization species are formed closer to the central source and are therefore associated with higher velocities. We will discuss the physical implications of the inferred fit parameters more in Section \ref{sec:discussion_state_characteristics}.

Our \hst~spectra show no indication of wind-formed UV features in the hard state, either in the form of blue-shifted absorption or P-Cygni profiles, as can be seen in Figure \ref{fig:MAXI_J1820_line_fits}. There is also no hint of such features in the \astrosat~soft state observations, but the lower signal-to-noise and resolution of those observations would prevent the detection of all but the strongest such features.

\begin{table*}
\caption{Attributes of the most prominent resonance line profiles in the UV region for all of our epochs. Specifically, we take a closer look at the usual suspects: the \ion{N}{v} $\lambda 1240$\AA, \ion{Si}{iv} $\lambda 1400$\AA, \ion{C}{iv} $\lambda 1550$\AA, \ion{He}{ii} $\lambda 1640$\AA~and \ion{Mg}{ii} $\lambda 2800$\AA~lines. The table below summarizes various line properties, derived from Gaussian fitting of the dereddened continuum-subtracted spectra. Key properties include the integrated fluxes (in $ \rm 10^{-12}\ erg\ s^{-1}\ cm^{-2}$) and EWs along with their associated errors. Additional parameters of the fitting such as the FWHM, the individual component velocity, the peak-to-peak separation (all in $\kms$), the doublet amplitude ratio as well as the ratio between their blue/red component are also shown. The quoted errors correspond to 1$\sigma$ confidence level.} \label{tab: line_profiles}
\begin{threeparttable}
\begin{tabular*}{\textwidth}{@{\extracolsep{\fill}} lcccccccc}
\hline
\hline
Line & $\rm F_{\lambda}$ & $\rm F_{\lambda, downgraded} $ \tnote{$\dagger$} & EW  & FWHM & $\sigma$ & $\rm v_{pp}$ & Doublet & Blue-to-red \\
& (cgs) & (cgs) & (\AA) & ($\kms$) & ($\kms$) & ($\kms$) & amplitude ratio & amplitude ratio \\
\hline
\hline
\multicolumn{9}{c}{\textbf{HST:HS1}} \\ 
\hline
\hline
\ion{N}{v} & 32.62 $\pm$ 0.92 & $-$ & 16.03 $\pm$ 0.80 & 1818 & 365 & 923 & 1.14 & 0.88 \\
\ion{Si}{iv} & 6.50 $\pm$ 0.22 & 6.35 $\pm$ 0.33 & 4.00 $\pm$ 0.16 & $-^{\ast}$ & 332 & 867 & 1.02 & 0.98  \\
\ion{C}{iv} & 8.71 $\pm$ 0.26 & 6.41 $\pm$ 0.15 & 7.53 $\pm$ 0.28 & 1773 & 446 & 869 & 1.27 & 0.43  \\ 
\ion{He}{ii} & 5.43 $\pm$ 0.18 & 5.37 $\pm$ 0.08 & 5.11 $\pm$ 0.19 & 1515 & 335 & 784 & 0.94 & $-$  \\
\ion{Mg}{ii} & 1.44 $\pm$ 0.04 & $-$ & 3.59 $\pm$ 0.12 & 1359 & 273 & 687 & 2.24 & 1.0 \\
\hline
\hline
\multicolumn{9}{c}{\textbf{HST:HS2}} \\ 
\hline  
\hline
\ion{N}{v} & 13.08 $\pm$ 0.47 & $-$ & 21.21 $\pm$ 1.33 & 2354 & 651 & 1366 & 3.48 & 1.07  \\
\ion{Si}{iv} & 2.14 $\pm$ 0.04 & $-$ & 3.98 $\pm$ 0.08 & $-$ & 466 & 1018 & 1.29 & 0.94  \\
\ion{C}{iv} & 2.62 $\pm$ 0.03 & $-$ & 6.78 $\pm$ 0.09 & 2231 & 460 & 983 & 0.52 & 0.46   \\ 
\ion{He}{ii} & 1.92 $\pm$ 0.04 & $-$  & 5.66 $\pm$ 0.14 & 1482 & 390 & 998 & 0.48 & $-$ \\
\ion{Mg}{ii} & 0.43 $\pm$ 0.01 & $-$  & 2.95 $\pm$ 0.07 & 1390 & 273 & 882 & 0.33 & 1.0 \\
\hline
\hline
\multicolumn{9}{c}{\textbf{AstroSat:HS}} \\
\hline
\hline
\ion{Si}{iv} & 7.23 $\pm$ 0.72 & $-$ & 4.66 $\pm$ 0.48 & $-$ & $-$ & $-$ & $-$ & $-$ \\
\ion{C}{iv} & 6.00 $\pm$ 0.43 & $-$ & 4.98 $\pm$ 0.41 & $-$ & $-$ & $-$ & $-$ & $-$  \\ 
\ion{He}{ii} & 4.84 $\pm$ 0.39 & $-$ & 4.5 $\pm$ 0.39 & $-$ & $-$ & $-$ & $-$ & $-$ \\
\hline
\hline
\multicolumn{9}{c}{\textbf{AstroSat:SS}} \\
\hline
\hline
\ion{Si}{iv} & 3.57 $\pm$ 0.41 & $-$ & 6.28 $\pm$ 0.76 & $-$ & $-$ & $-$ & $-$ & $-$ \\
\ion{C}{iv} & 3.59 $\pm$ 0.39 & $-$ & 8.65 $\pm$ 1.01 & $-$ & $-$ & $-$ & $-$ & $-$ \\ 
\ion{He}{ii} & 3.17 $\pm$ 0.74 & $-$ & 9.13 $\pm$ 2.32 & $-$ & $-$ & $-$ & $-$ & $-$ \\
\hline
\hline
\multicolumn{9}{c}{\textbf{X-shooter:HS \tnote{$\ddagger$}}} \\
\hline
\hline
$\rm H{\alpha}$ & 0.247 $\pm$ 0.003 & $-$ & 7.03 $\pm$ 0.08 & 649 & 248 & 732 & 2.14 & $-$ \\
\hline
\hline
\end{tabular*}
\begin{tablenotes} [para, flushleft]
\textbf{Notes:}\item[$\dagger$] Line flux measurements from the downgraded dereddened continuum-subtracted spectra for an immediate comparison to the AstroSat:SS estimates. This comparison is imminently seen in Figure \ref{fig:MAXI1820_line_flux_evolution} where we follow the line flux evolution of the source throughout the three epochs of the outburst. The respective uncertainties are estimated through Monte Carlo simulations. \\
\item[$\ast$] The big separation ($\approx$9\AA) between the components of the \ion{Si}{iv} $\mathrm \lambda1400$\AA~line results in doubling the apparent width and hence the FWHM of its integrated profile. \\ 
\item[$\ddagger$] The \textit{VLT}/X-shooter observation is taken  few days before the HST:HS1 epoch, on 2018 March 22nd (or in MJD:58199.32). 
\end{tablenotes}
\end{threeparttable}
\end{table*}

\section{Irradiated disc modelling}  \label{sec:
Irradiated_disc_modelling}

X-ray irradiation is a crucial, but poorly understood, aspect of the BHXT eruptions. The reprocessing of X-ray photons can dominate over viscous heating in the outer disc, thereby generating the UV and optical light emitted in these regions \citep[e.g.][]{vanparadijs94, vanparadijs96, king98, dubus01, dubus99, tetarenko18b}. However, many aspects of this picture remain unclear, such as the dependence of the reprocessing efficiency on the irradiating SED. 

As noted above (see Section \ref{sec:spectroscopic_overview} and Figure \ref{fig:MAXI_J1820_spectral_evolution}), the UV continuum shape of \source~seems to pose a challenge in this context. Given the high X-ray luminosity in each of our three epochs, irradiation might be expected to dominate over viscous energy release in the UV-producing disc regions. Yet the observed continuum shape is better approximated by that associated with a viscously-dominated disc ($\mathrm F_{\lambda, \rm visc} \propto \lambda^{-7/3}$) than that associated with an irradiation-dominated disc ($\mathrm F_{\lambda, \rm irr} \propto \lambda^{-1}$). Similar behaviour has also been observed in other BHXTs, specifically Nova Muscae 1991 \citep{cheng92}, XTE J1859+226 \citep{hynes02} and A 0620-00 \citep{hynes05}. To investigate this further, we have tried to model the UV continuum more quantitatively with a model that includes both viscous and irradiation-driven heating.

Below, we first introduce our model and the irradiation contribution to the energy balance in the disc (Section \ref{sec:irradiated_model}). We then used the HST:HS2 spectrum as a test bed to examine whether a simple irradiated disc model can replicate the flux level and spectral shape of the observed UV SED (Section~\ref{sec:disc_SED_modelling}).

\subsection{Description of the model} \label{sec:irradiated_model}

Our model describes the disc as a collection of concentric circular annuli, each of which is characterized by an effective temperature $\rm T_{eff}(R)$. The disc extends from ISCO, $ \rm R = R_{ISCO}$, to an outer radius, $\rm R = R_{disc}$. The effective temperature is set by the requirement that the rate at which an annulus radiates energy away, $\rm \sigma T_{eff}^4$, must balance the rate at which energy is deposited into it by viscous dissipation and irradiation. 

The viscous heating rate can be written as 
\begin{equation}
\rm \sigma T_{visc}^4 = \frac{3 G M_{BH} \dot{M}_{acc}}{8\pi R^3} \left[ 1 - \left(\frac{R_{ISCO}}{R}\right)^{1/2}\right],
\end{equation}
where $\rm M_{BH}$ is the mass of the BH, and $\rm \dot{M}_{acc}$ is the accretion rate. The heating rate due to irradiation can be modelled as

\begin{equation}
\rm \sigma T_{irr}^4 = \left(\frac{L_{irr}}{4\pi R^2}\right) 
    \left(\frac{H}{R}\right) \gamma 
    \left(1-A\right),
\end{equation}

where $\rm L_{irr}$ is the irradiating luminosity (assumed to originate from a central point source), and $\rm A$ is the albedo (so that $\rm 1-A$ is the fraction of the light incident on the annulus that is absorbed). The quantity $\rm H/R$ is the aspect ratio of the disc, which can be shown \citep{frank02} to scale as 

\begin{equation}
  \rm  \frac{H}{R} = \left(\frac{H}{R}\right)_{R_{disc}} \left(\frac{R}{R_{disc}}\right)^\gamma.
\end{equation}

Here, the power-law index is usually taken to be $\rm \gamma = 1/8$ in the absence of irradiation, and $\rm \gamma = 2/7$ if irradiation dominates the heating rate \citep[c.f.][]{frank02}. Strictly speaking, $\rm \gamma$ is therefore a function of radius, but we neglect this here and simply set $\rm \gamma = 2/7$ everywhere. This approximation means that we will slightly overestimate the influence of irradiation, but only in disc regions where irradiation is relatively unimportant anyway. 

Putting all of this together, the effective temperature of the disc can be calculated by requiring that total heating should be matched by radiative cooling, i.e.
\begin{equation}
  \rm  \sigma T_{eff}^4 = \sigma T_{visc}^4 + \sigma T_{irr}^4.
\end{equation}

In order to calculate the spectrum of the disc, we assume that each annulus radiates as a modified blackbody, 
\begin{equation}
 \rm   B_{\nu, mod}(f,T_{eff}) = \frac{2 h \nu^3}{f^4 c^2 
    \left[e^{\frac{h\nu}{f k T_{eff}}} - 1\right]}.
\end{equation}
Here, $\rm f$ is the so-called ``spectral hardening factor'' \citep{shimura95}, which approximately corrects for the effects of Compton scattering in the disc atmosphere. This factor is not actually a constant, but rather is a function of temperature, surface density and radius (and, for BHs, spin parameter). In our SED model, we parameterize $\rm f$ using the analytical fitting function provided by \citet{davis19}. In calculating the required surface density, we take into account the relevant relativistic correction factors \citep{novikov73, davis19}. In our specific application of the model to UV observations here, the effects of spectral hardening are negligible.

Empirically, the SEDs of X-ray binaries exhibit at least one additional high-energy component, which is usually attributed to a hot, compact and optically thin "corona" located close to the accretor. We model this component, which usually dominates in the hard X-ray regime, as a simple power-law, 
\begin{equation}
\rm I_{\nu, cor} = I_{\nu_{0}, cor} \left(\frac{\nu}{\nu_{0}}\right)^{-(\alpha+2)}.
\end{equation}
With our definition of $\alpha$, we have $\rm I_{\lambda, cor} \propto \lambda^\alpha$, and we adopt $\rm \nu_0 = 3\times 10^{18}$~Hz (corresponding to $\lambda = 1$~\AA) as the reference frequency. However, this component is not relevant to our application.

\subsection{Disc SED modelling}   \label{sec:disc_SED_modelling}

Our primary focus is to assess if irradiation significantly impacts the outer accretion disc regions and can reproduce the shape and flux level of the far- and near-UV SED. To achieve this, we opt to model the spectrum prior to the state transition (HST:HS2) as it provides us a unique opportunity for a direct comparison of our model, which is close in time and luminosity, with the one employed by \citet{koljonen23}, henceforth referred to as the "reference model". We note that in our performed modelling, we fix $\rm L_x = 1.35 \times 10^{38} \lum$ \citep{koljonen23}.

We first try to find the optimal model that describes our dereddened UV SED.  We allow the mass accretion rate ($\rm \dot M_{acc}$), outer disc radius ($\rm R_{out}$) and albedo ($\rm A_{out}$) to vary, taking into account both irradiation heating and viscous dissipation. It is important to mention that we cannot avoid the mathematical degeneracy between albedo and scale height, so we opt to fix the latter for simplicity. We assume a Schwarzschild (non-spinning) BH \citep{guan21, zhao21}, scale height $\rm H/R$=0.1 while the orbital characteristics (d, $\rm M_{BH}$, i, q) of the system are already determined \citep{atri20, torres20}. In particular, we adopt $\rm M_{BH} = 8.5 M_{\odot}$ and $\rm i=65\degree$ in our modelling, while all the distances are measured with respect to the gravitational radius, $\rm R_G = GM_{BH}/c^2$. 

\begin{figure}
\centering
\includegraphics[width=0.49\textwidth]{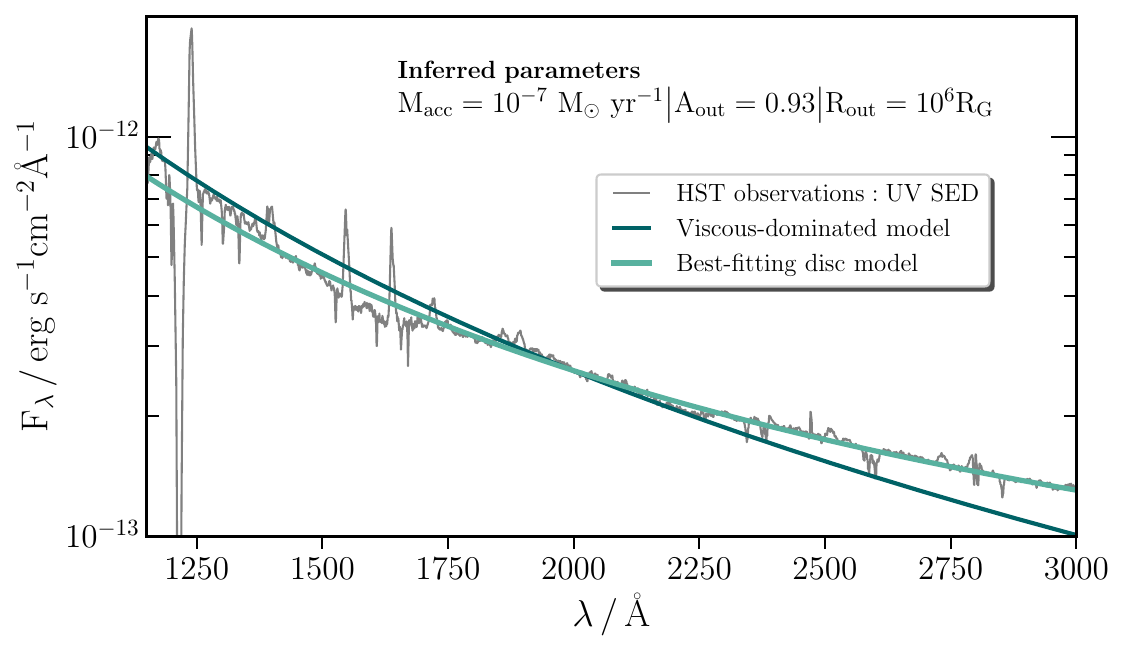}
\includegraphics[width=0.49\textwidth, height=0.53\textheight, keepaspectratio]{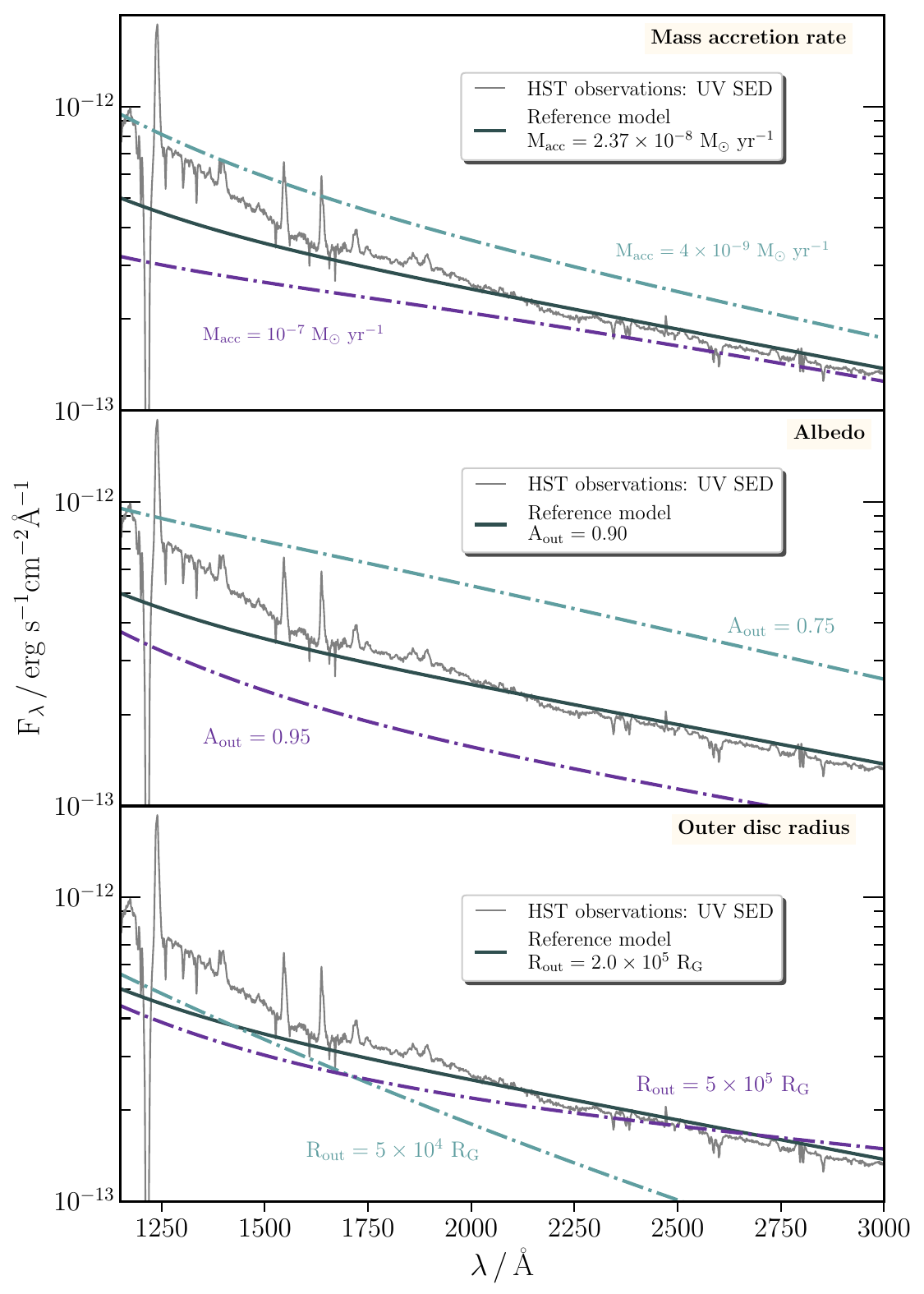}
\caption{Modelling of the UV SED of \source~during the epoch preceding the hard-to-soft state transition (HST:HS2). \textit{Upper panel}: The best-fitting model for our SED is very close to a pure viscous-dominated disc. This model results in a set of unrealistic values for parameters such as $\rm M_{acc}, A_{out}, R_{out}$, making it unsuitable for describing the system. \textit{Lower panel}: Parameter sensitivity analysis in our modelling to evaluate the dependency and impact of the considered disc factors on the final output. Each panel corresponds to one of these factors, where we explore a plausible range of their parameter space (dashdot lines), overlaid on the observed HST:HS2 UV spectrum (grey colour) of \source. The reference disc UV SED, as modelled by \citet{koljonen23}, is included in dark grey across all panels for context.}
\label{fig:MAXI1820_irradiation_modelling}
\end{figure}

The best-fitting model SED is shown in Figure \ref{fig:MAXI1820_irradiation_modelling} (upper panel). As already noted above, the observed spectral shape of the UV continuum is close to that of a standard accretion disc without any irradiation. In line with this, the UV emitting regions in our best-fitting model are \textit{viscously-dominated}, with very little, contribution from irradiation. However, the parameters of our optimal model $-$ $\rm \dot{M_{acc}} = 10^{-7}\ \mdot \simeq 0.5 \dot{{M}_{Edd}}$, $\rm R_{out} = 10^6 R_G$, $\rm A_{out}=0.93$ $-$ are physically implausible. In particular, the radius of the outer disc is much greater than the radius of the \textit{ tidal disc} of the BH accretor, as $\rm R_{out} \approx 4\times R_{tidal}$. For reference, the tidal radius is estimated as $\rm R_{tidal} = 10^{5.4} R_G$.

Can a physically more plausible irradiated disc model still reproduce the SED acceptably? Naively, for any reasonable values of the disc parameters ($\rm \dot M_{acc}, R_{out}, A_{out}$), irradiation should play a crucial role. To address this question more directly, we construct a small grid of models that allows us to understand how changes in these parameters affect the overall SED (and the irradiation contribution, specifically). Our grid is roughly centered on the "reference" disc SED model of \citet{koljonen23}. This model is derived by fitting the X-ray-optical SED of \source~at the soft state, days after the hard/soft state transition. The corresponding reference parameters are $\rm \dot{M}_{acc} = 2.37 \times 10^{-8}\ \mdot$, $\rm R_{out} = 10^{5.3} R_G$ and $\rm A_{out}$ = 0.9. The attempted comparison is justified as the two epochs are close enough in time that we do not expect the values of the considered parameters to change significantly.

Figure \ref{fig:MAXI1820_irradiation_modelling} (lower triplet of panels) shows the reference and test models that make up our small grid overlaid on the second hard-state UV SED. The reference model is clearly too red and underestimates the far-UV flux level. Moreover, even though our test models confirm that reasonable changes to the model parameters can certainly modify the UV continuum, they also suggest that such changes cannot simultaneously match the brightness and shape of the observed SED. We will discuss the implications of this result in Section \ref{sec:discussion_irradiation}.

\section{Timing analysis}  \label{sec:timing_analysis}

In this section, we search for and analyze the variability in our UV data set. This is particularly interesting because the emission produced in the \textit{outer} accretion disc is likely to peak in this waveband. To this end, we first construct wavelength-integrated light curves from our time-resolved observations, and then the corresponding PDS, where the variability amplitude (power) is expressed as a function of frequency. 

\begin{figure*}
\includegraphics[width=\textwidth, height=0.91\textheight]{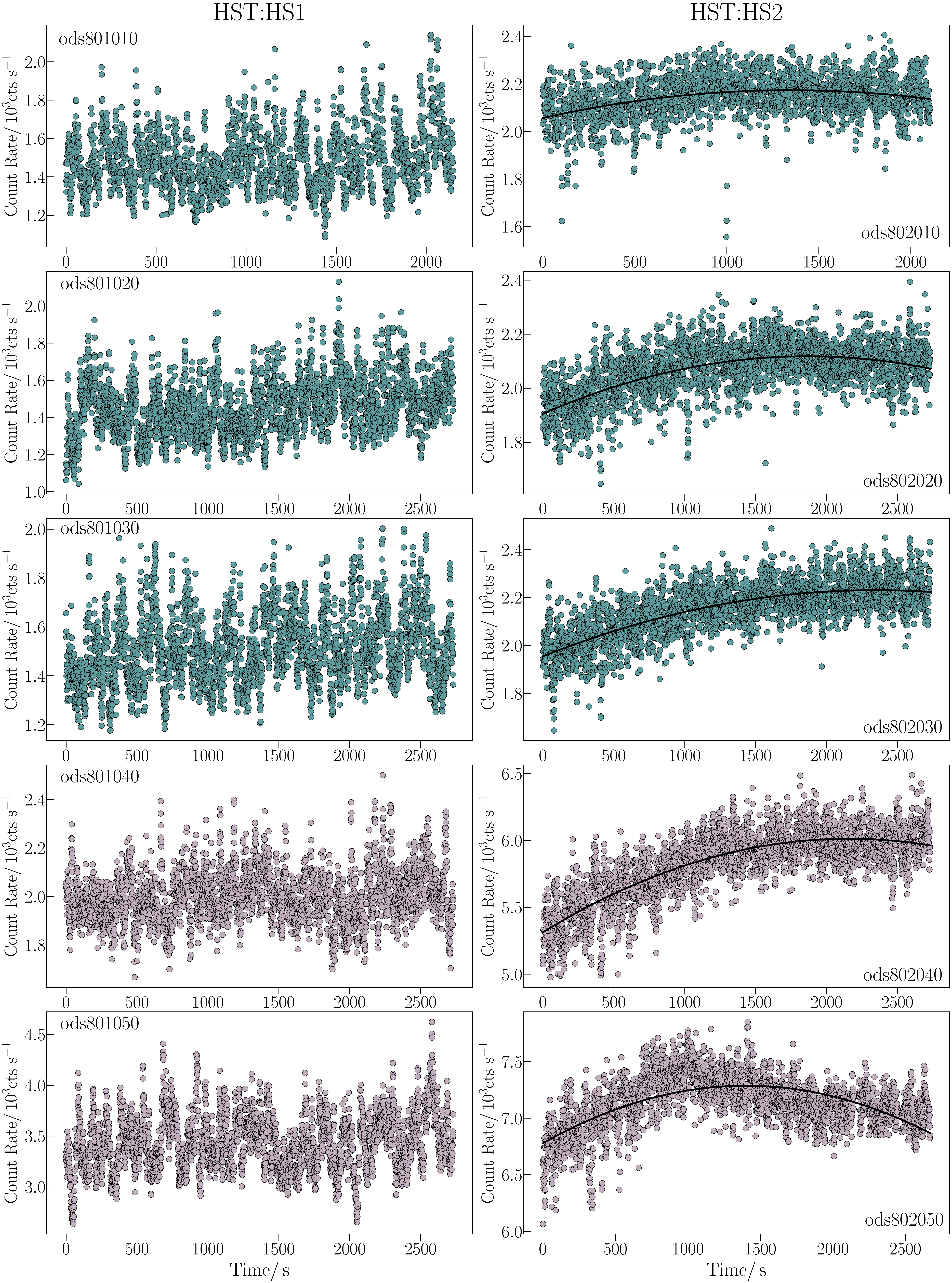}
\captionsetup{width=\textwidth}
\caption{Far- and near-UV light curves of \source, extracted at one-second resolution, for both of our hard state observations, taken by \hst. As noted, the left column showcases light curves of the first luminous hard state (HST:HS1) while the right column features the corresponding light curves of the second hard state (HST:HS2). For our HST:HS2 observations, we overlaid our polynomial fit, correcting for the presented turnover. The two colours are utilized to distinguish between the far- (teal) and near-UV observations (dusty pink). For completion, we have also noted the obsID of the considered observations at each of the panels.}
\label{fig:maxi1820_overall_lc}
\end{figure*}

The light curves are generated using the \textsc{lightcurve} package.\footnote{The original code is available at {\small \url{https://github.com/justincely/lightcurve}}, but we have adapted the code to suit our requirements.} They are both background-subtracted and corrected for buffer dumps. An overview of the light curves for all epochs can be seen in Figure \ref{fig:maxi1820_overall_lc}. Rapid, aperiodic variability is apparent throughout the observations. More specifically, the far-UV light curves obtained during the HST:HS1 observations clearly exhibit flaring activity with a fractional rms amplitude of $\simeq 10\%$. The corresponding flaring amplitude in the near-UV is $\simeq 7\%$.

Rather suspiciously, all of the light curves obtained from the HST:HS2 visit in Figure \ref{fig:maxi1820_overall_lc} exhibit a slow rise in count rate at the beginning of each orbit. Following discussions with the STScI office and a detailed inspection of the cross-dispersion profiles during this visit, we tentatively attribute these variations to instrumental focus changes. We therefore opt to filter out these slow variations by fitting a second-order polynomial to the respective light curves and then normalizing them to this fit. The polynomial fits are shown overlaid on the raw HST:HS2 light curves in Figure \ref{fig:maxi1820_overall_lc}. We then measure the fractional rms amplitudes of the short-term variability in each orbit, which is $\simeq 4\%$ and $\simeq 3\%$ in the far- and near-UV, respectively.  

Next, we construct the PDS following the methodology by \cite{vaughan05}. We define the periodogram for a light curve with K points and $\mathrm{\Delta T}$ sampling rate as the modulus-squared of the discrete Fourier transform (DFT), $ \mathrm{X(f_{j})}$,  

\begin{equation}
    \rm I(f_j) = \frac{2 \Delta T}{<x>^2 N} |X_j|^2,
\end{equation}
evaluated at the Fourier frequencies $ \rm f_j = j/K\Delta T$ with j=1,2,..., K/2. Its normalization is in $\rm (rms/mean)^2 Hz^{-1}$, expressing the power in fractional units \citep{vanderklis89, vanderklis97, vaughan03, vaughan05}.

\begin{figure*}
\includegraphics[width=\textwidth]{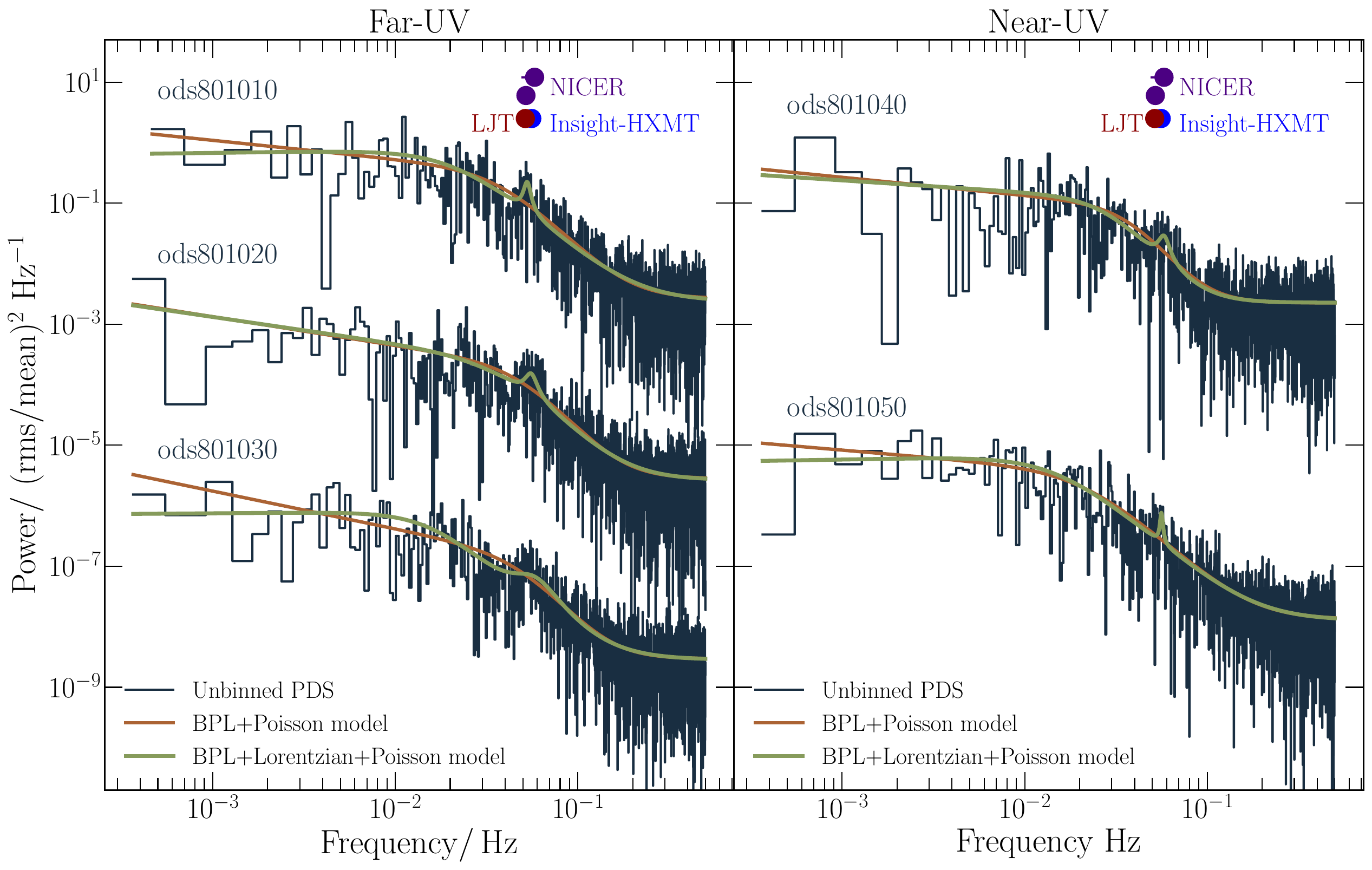}
\caption{The PDS for both the far- (left) and near-UV (right) wavelengths during our HST:HS1 epoch are presented. The different solid lines represent the best-fitting models, incorporating the Lorentzian contribution. Tentative evidence for the presence of a LF-QPO is observed in the far-UV, which is absent in the near-UV observations. For reference, characteristic centroid QPO frequencies, estimated in the X-ray (NICER: \citealt{stiele2020}, Insight-HMXT/HE: \citealt{mao22}) and optical \citep[LJT/YFOSC:][]{yu2018a, mao22} bands, close to our observation time, are displayed. }
\label{fig:HS1_UV_PDS_panel}
\end{figure*}

One of our aims here is to examine our datasets for the presence of any significant sharp features, e.g., low-frequency quasi-periodic oscillations (LF-QPOs), as seen in X-ray and optical studies \citep[e.g.][]{stiele2020, mudambi20, paice21, thomas22, mao22}. Overall, our UV PDS exhibit a pattern consistent with that seen in other sources and at other wavelengths. The "continuum" can be aptly described by a single-bend power-law with two distinct power indices, delineated by the break frequency. At the highest frequencies, the power spectra are governed by Poisson noise. Finally, especially during our HST:HS1 observations, we do observe a weak yet discernible signature of a QPO close to the break frequency, within the frequency range dominated by red noise, and at a frequency that is similar to QPOs detected in other bands.

We decided to take a closer look at the power spectra in which we tentatively detect a peak. Specifically, to determine the statistical significance of these detections, we fit each observed PDS with two models, one with and one without a QPO component. The intrinsic "continuum" in both models is described as a broken (or bending) power-law, while the QPO is approximated as a Lorentzian of a centroid frequency, $\rm v_c$, and FWHM width, $ \rm W_q$. 

The full PDS model that includes the QPO is given by \citet{summons07}
\begin{equation}
  \rm  P(\nu) = \frac{A_{BPL}\nu^{\alpha_L}}{1+(\frac{\nu}{\nu_{B}})^{\alpha_L-\alpha_H}} + \frac{B_{QPO}Q\nu_c}{\nu_c^2+4Q^2(\nu_c-\nu)^2} + C_{PS}.
\end{equation}
Here, the first term describes the broken power-law (BPL) continuum with $\rm \nu_{B}$ the break frequency, and $\rm \alpha_L$, $\rm \alpha_H$, the low and high frequency indices. The second term accounts for the presence of the QPO with the parameter Q, known as the quality factor (measure of coherence), aptly connected to the centroid frequency as $\rm v_c/W_q$. The third term, $\rm C_{PS}$, describes the variability associated with Poisson noise. The parameters $\rm A_{BPL}$ and $\rm B_{QPO}$ serve as normalization constants for the individual components. In the model without a QPO, the second term is simply set to zero. 

\begin{table*}
\caption{Best-fitting model parameters of the considered models for the HST:HS1 PDS both at the far- and near-UV range.} \label{tab:PDS_modelling}
\begin{adjustbox}{width=\textwidth, keepaspectratio}
\begin{tabular*}{1.2\textwidth}{@{\extracolsep{\fill}} ccccccccccc}
\hline
\hline
\multicolumn{11}{c}{\textbf{HS1: far-UV @1425 \AA}} \\ 
\hline
ObsID & $\rm logA_{BPL} ^{(a)}$ & $\rm \alpha_L ^{(b)}$ & $ \rm \alpha_H ^{(c)}$ & $\rm \nu_B ^{(d)}$ & $\rm logB_{QPO} ^{(e)}$ & $ \rm logQ ^{(f)}$ & $\rm \nu_c ^{(g)}$ & $\rm C_{PS} ^{(h)}$ &  KS-Test $\rm ^{(i)}$ & $\rm \Delta S ^{(j)}$ \\
& & & & (mHz) & & & (mHz) & &  & \\
\hline 
ods801010 & -0.044 $\pm$ 0.53 & 0.042 $\pm$ 0.23 & -2.52 $\pm$ 0.19 & 20.93 $\pm$ 5.42 & -3.10 $\pm$ 0.25 & 1.01 $\pm$ 0.49 & 52.79 $\pm$ 1.63 & (2.38 $\pm$ 0.21) $\times 10^{-3}$ & 0.74 \\
& -0.9 $\pm$ 0.35 & -0.31 $\pm$ 0.16 & -3.09 $\pm$ 0.26 & 38.60 $\pm$ 7.53 & $-$ & $-$ & $-$ & (2.63 $\pm$ 0.19) $\times 10^{-3}$ & 0.62 & 0.044 \\
\hline
ods801020 & -1.20 $\pm$ 0.40 & -0.44 $\pm$ 0.18 & -2.95 $\pm$ 0.35 & 37.44 $\pm$ 11.25 & -3.07 $\pm$ 0.30 & 0.77 $\pm$ 0.47 & 55.34 $\pm$ 2.48 & (2.67 $\pm$ 0.19) $10^{-3}$ & 0.97 \\
& -1.30 $\pm$ 0.25 & -0.475 $\pm$ 0.13 & -3.47 $\pm$ 0.27 & 49.85 $\pm$ 7.24 & $-$ & $-$ & $-$ & (2.81 $\pm$ 0.15) $\times 10^{-3}$ & 0.86 & 0.014 \\
\hline
ods801030 & -0.057 $\pm$ 0.58 & 0.024 $\pm$ 0.24 & -2.95 $\pm$ 0.98 & 16.00 $\pm$ 4.55 & -2.75 $\pm$ 0.24  & 0.18 $\pm$ 0.19 & 54.65 $\pm$ 5.82 & (2.84 $\pm$ 0.14) $10^{-3}$ & 0.46 \\
& -1.63 $\pm$ 0.28 & -0.62 $\pm$ 0.14 & -3.56 $\pm$ 0.32 & 49.31 $\pm$ 8.48 & $-$ & $-$ & $-$ & (2.96 $\pm$ 0.13) $\times 10^{-3}$ & 0.61 & 0.044 \\
\hline
\hline
\multicolumn{11}{c}{\textbf{HS1: near-UV @1978 \AA}} \\ 
\hline
ods801040 & -1.22 $\pm$ 0.38 & -0.20 $\pm$ 0.18 & -3.63 $\pm$ 0.55 & 28.88 $\pm$ 5.11 & -3.80 $\pm$ 0.28  & 0.82 $\pm$ 0.40 & 57.81 $\pm$ 2.11 & (2.26 $\pm$ 0.083) $\times 10^{-3}$ & 0.48 \\
& -1.48 $\pm$ 0.29 & -0.30 $\pm$ 0.14 & -3.88 $\pm$ 0.41 & 37.40 $\pm$ 5.10 & $-$ & $-$ & $-$ & (2.25 $\pm$ 0.080) $\times 10^{-3}$ & 0.39 & 0.0067  \\
\hline
\hline
\multicolumn{11}{c}{\textbf{HS1: near-UV @2707 \AA}} \\ 
\hline
ods801050 & -0.075 $\pm$ 0.055 & 0.054 $\pm$ 0.22 & -2.49 $\pm$ 0.15 & 14.65 $\pm$ 3.29 & -3.97 $\pm$ 0.96 & 1.44 $\pm$ 3.00 & 56.13 $\pm$ 0.90 & (1.28 $\pm$ 0.087) $\times 10^{-3}$ & 0.74 \\
& -0.85 $\pm$ 0.45 & -0.256 $\pm$ 0.19 & -2.70 $\pm$ 0.17 & 21.64 $\pm$ 4.69 & $-$ & $-$ & $-$ & (1.33 $\pm$ 0.082) $\times 10^{-3}$ & 0.83 & 0.069  \\
\hline
\hline
\end{tabular*}
\end{adjustbox}
\justify
\textbf{Notes:} All the uncertainties are quoted to 1$\sigma$ confidence level. The second line of each observation corresponds to our modelling without the Lorentzian contribution. $\rm ^{(a)}$ BPL normalization constant. $\rm ^{(b)}$ Low-frequency (before the break) power-law index. $\rm ^{(c)}$ High-frequency power-law index. $\rm ^{(d)}$ Break frequency. $\rm ^{(e)}$ Normalization constant of the Lorentzian, describing the QPO. $\rm ^{(f)}$ Quality factor defined as $\rm v_c/W_q$, where $\rm W_q$ the FWHM of the Lorentzian. $\rm ^{(g)}$ Centroid frequency of the QPO peak. $\rm ^{(h)}$ Poisson constant. $\rm ^{(i)}$ Goodness-of-fit p-value obtained via the KS-test. $\rm ^{(j)}$ Statistical significance of the QPO obtained through the LRT test.
\end{table*}

Following \citet{vaughan05}, we use the maximum likelihood estimation (MLE) method and the S-statistic in order to find the optimal parameters of the two models. All model parameters are allowed to vary. We assess the goodness-of-fit in both cases using the Kolmogorov-Smirnov (KS) test, comparing the data/model residual ratio with the theoretical $\rm \chi^2$ distribution of two degrees of freedom, $\rm \chi^2_2$, for unbinned periodograms. Further details \footnote{We compare the data/model residual ratio to a theoretical $\rm \chi^2_2$ distribution as the real and imaginary terms of a DFT are normally-distributed and their superposition follows a $\chi^2_2$ variable.} can be found in \citet{vaughan05} and references therein. The uncertainties in our best-fit parameters are estimated by Monte Carlo simulations. For each dataset, we generate 1000 mock exponentially-distributed PDS and fit them in the same manner as described earlier. This iterative process provides us with the best-fit parameters and S-statistic for each mock dataset, allowing us to estimate errors at 1$\sigma$ confidence level. The best-fitting parameters and their uncertainties, as well as details of our fitting results, are summarized in Table \ref{tab:PDS_modelling}.

\begin{figure}
\includegraphics[width=0.5\textwidth]{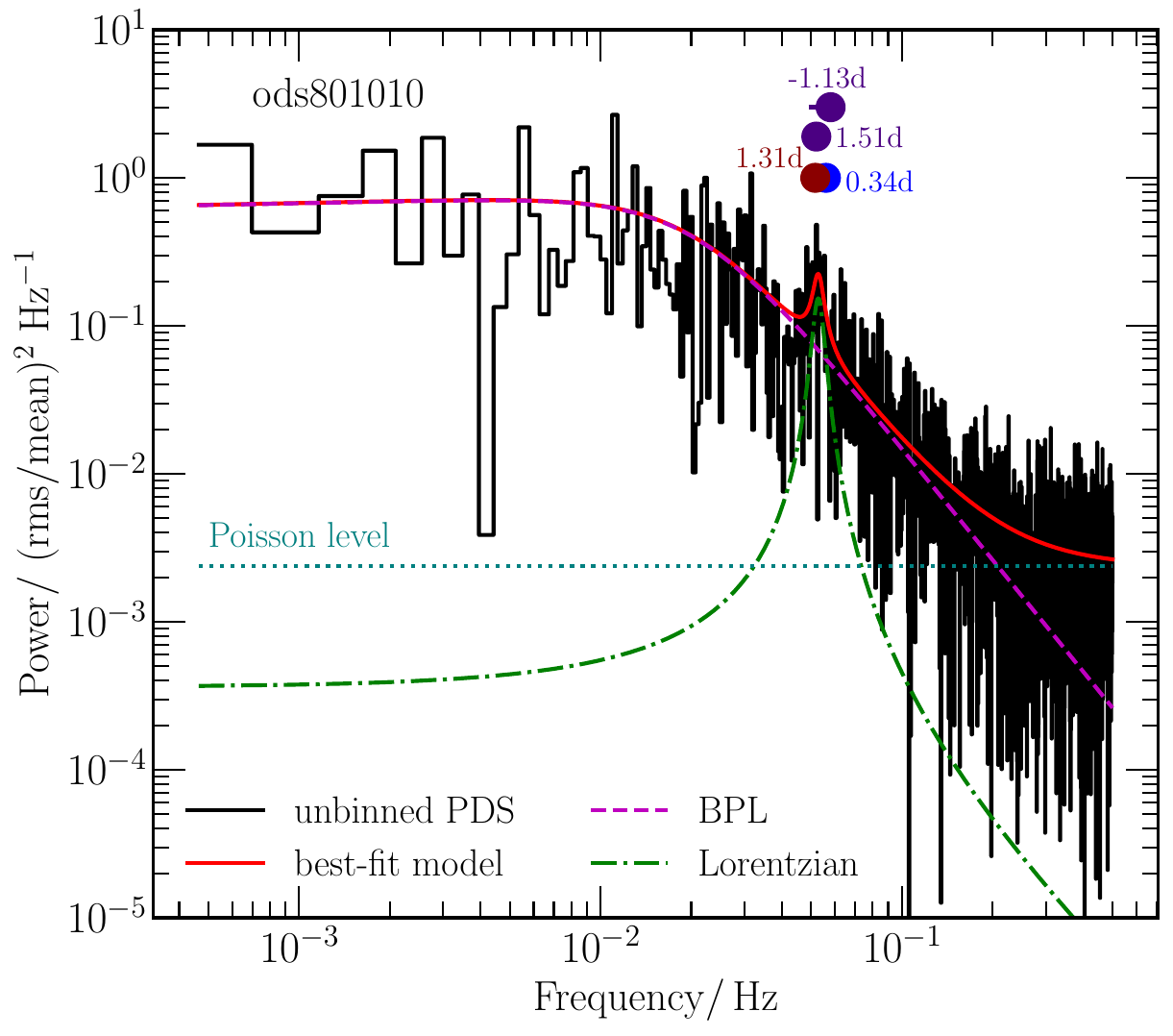}
\caption{Example of a far-UV PDS of our source during our HST:HS1 observations, showing a weak signature of a QPO. The solid red line represents the overall best-fit model while the dashed and dash-dotted lines highlight the two main model contributions (BPL: magenta, Lorentzian: green). The level of Poison noise is seen as a teal dotted line. The X-ray and optical QPO frequencies,  mentioned in Figure \ref{fig:HS1_UV_PDS_panel}, offer valuable context for comparison. }
\label{fig:1820_HS1_PDS_III_component_model}
\end{figure}

Figure \ref{fig:HS1_UV_PDS_panel} presents the different PDS for both the far- and near-UV regions, where we have also superimposed both the QPO and non-QPO models. In Figure \ref{fig:1820_HS1_PDS_III_component_model}, we additionally illustrate the contributions of the different components. In both cases, we indicate the estimated centroid frequencies in both the X-ray \citep{stiele2020, mao22} and optical \citep{mao22} wavelengths, with respect to the time of our UV observations. Taken separately, both models provide statistically acceptable fits to the data. 

In order to test if the QPOs are significantly detected, we once again follow \citet{vaughan05} in estimating their statistical significance. Specifically, we use the likelihood ratio test (LRT) to check if the additional freedom associated with the QPO model is warranted by the data. The test statistic here is

\begin{equation}
    \rm \Delta S = S_1 - S_2 = -2ln[\mathcal L_1/ \mathcal L_2],
\end{equation}

where indices 1 and 2 represent the likelihoods and S-statistics for the simpler and complex models, respectively. If the models are nested, and the simpler model is correct, the variable $\rm \Delta S$ is $ \rm \chi^2_{\nu}-$distributed, where $\rm \nu$ is the number of additional free parameters. In our case, $\rm \Delta S$ therefore follows a $\rm \chi^2_{3}$ distribution under the null hypothesis that the no-QPO model is correct (since the QPO term requires three additional parameters). The p-values associated with this null hypothesis turn out to be 0.044, 0.014, 0.044 for the far-UV and 0.0067, 0.069 for the near-UV range. These values correspond to 2-3$\sigma$ detections of the QPOs $-$ suggestive, but not definitive. Given that the locations of the putative UV QPOs are consistent with those seen in other bands, we tend to think that they are real and worthy of consideration.

\begin{figure}
\includegraphics[width=0.5\textwidth]{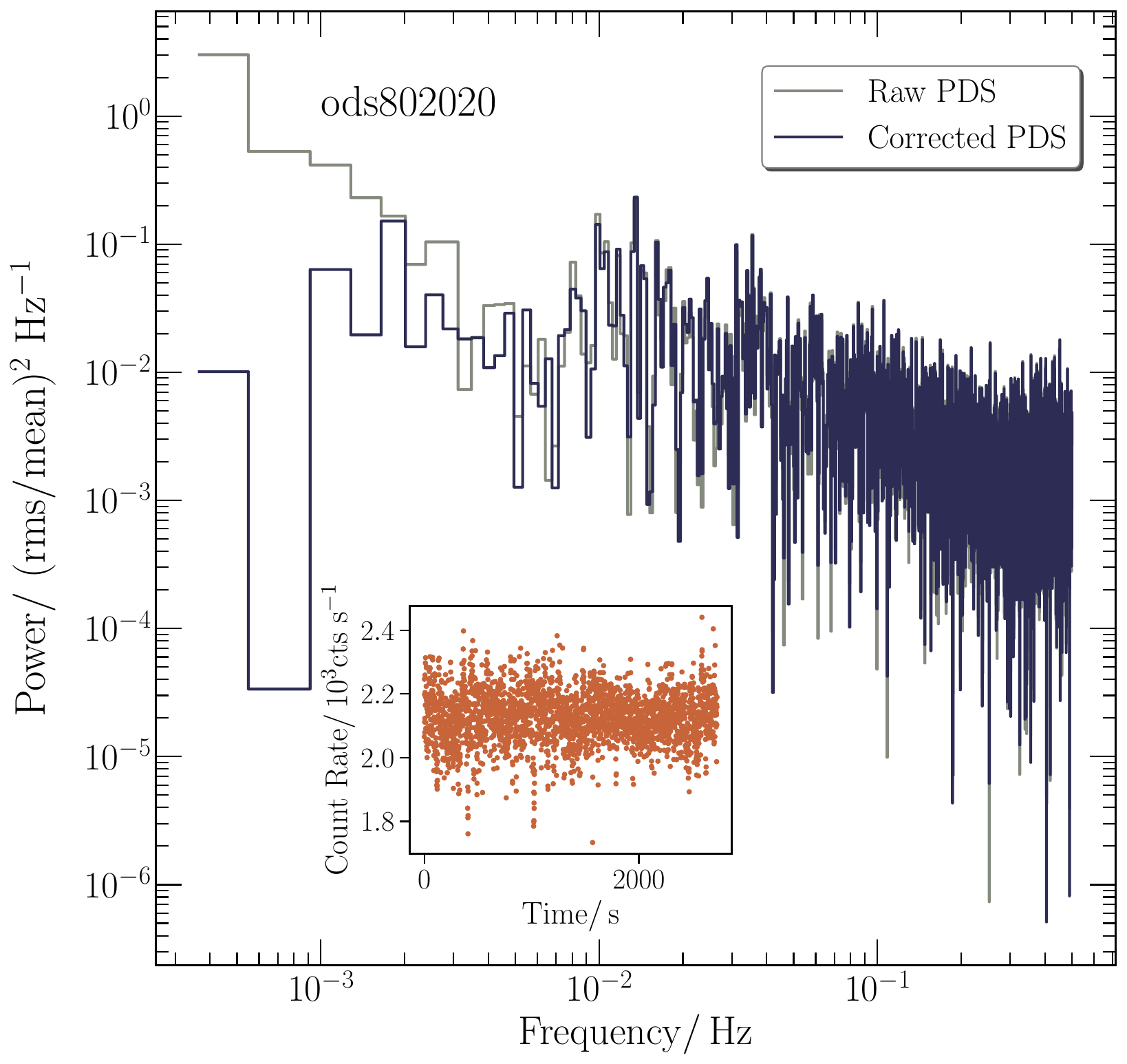}
\caption{Representative far-UV PDS of the HST:HS2 epoch of \source. We overlaid both the raw (grey) and corrected (dark blue) PDS to illustrate the difference prior and after the applied correction. The corresponding corrected light curve is placed as an inset.}
\label{fig:HST:HS2_PDS_comparison}
\end{figure}

We also constructed a PDS from both the raw and corrected light curves of our HST:HS2 observations. Notably, all the raw PDS exhibit an unusual excess at low frequencies, in line with our suspicion that the slow variations in these data sets are instrumental artefacts. This low-frequency excess disappears once the polynomial fit is used to "correct" the light curves. An illustrative example of this process is presented in Figure \ref{fig:HST:HS2_PDS_comparison}. There are no hints of QPO feature in HST:HS2, in any of the visits associated with this epoch. Given this, and also the inevitable uncertainty associated with our polynomial correction, we choose not to perform detailed PDS modelling for this epoch. 

\section{Discussion}  \label{sec:discussion}

\subsection{The spectral journey through the accretion states}  \label{sec:discussion_spectroscopy}

\subsubsection{Probing the accretion states}  \label{sec:discussion_state_characteristics}

Accretion states in BHXTs are usually defined in terms of their X-ray properties. Do their UV properties reflect the same phenomenological picture?

\source~provide us with a rare opportunity to investigate, for the first time in the UV band, the long-term spectral evolution of a BH binary as it transitioned from the hard to the soft state. Interestingly, we find that the UV behaviour remains consistent among both states: there are no distinct signatures that would signal the state transition or the different phases of the hard-state decay. This phenomenon may come as a surprise if we consider the case of the BHXT XTE J1859+226 \citep{hynes02}, where we do observe evolution of the spectrum during different stages of its outburst decay. The UV spectra exhibit the same Doppler-broadened, double-peaked line profiles, which are most likely emitted by an optically thin layer of the disc's atmosphere. 

\begin{figure}
\centering
\includegraphics[width=0.5\textwidth]{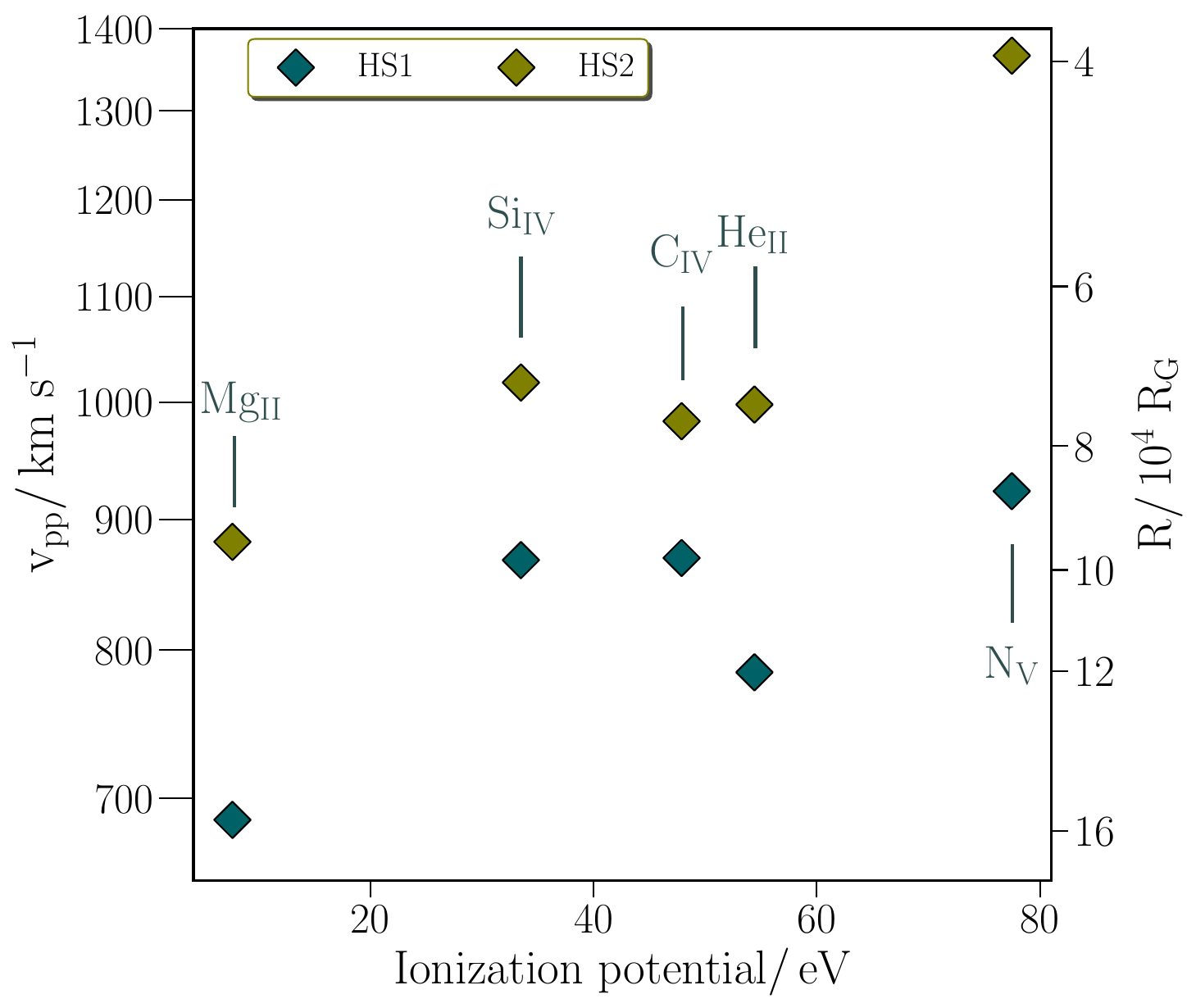}
\caption{The peak-to-peak velocities, $\rm v_{pp}$, of the main atomic species for both of our \hst~hard state epochs (HS1 $-$ in teal and HS2 $-$ in olive colour), as a function of their ionization potential. On a secondary axis, we depict the corresponding radii (in $\rm R_G$ units) where these species are formed in the disc. High-ionization emission lines are formed closer to the central object whereas low-ionization species are situated further out at the outer disc. For both epochs, the line-forming regions are located within the tidal radius.}
\label{fig:MAXI1820_p2p_veloc_radii_potential}
\end{figure}

Given the velocities determined from our modelling of the line profiles, we can estimate the location of the corresponding line-forming regions by assuming that the peak-to-peak separation of a given line corresponds to the Keplerian velocity near (the outer edge of) the disc region in which the line is produced. This line formation radius estimate is therefore given by
\begin{equation}
  \rm R = \pm \frac{GM_{BH}sini^2}{v_{pp}^2} \Rightarrow R = \frac{c^2 sini^2}{v_{pp}^2} [R_G],
\end{equation}
where the latter equation is expressed in terms of $\rm R_G$. This quantity is shown for each line in each \hst~epoch in Figure \ref{fig:MAXI1820_p2p_veloc_radii_potential}, where we also relate this to the ionization potential of the relevant species. As expected, the higher ionization lines are generated at smaller radii, where the temperatures are higher, while the lower ionization lines originate in the cooler parts of the disc further out.

\begin{figure}
\centering
\includegraphics[width=0.5\textwidth]{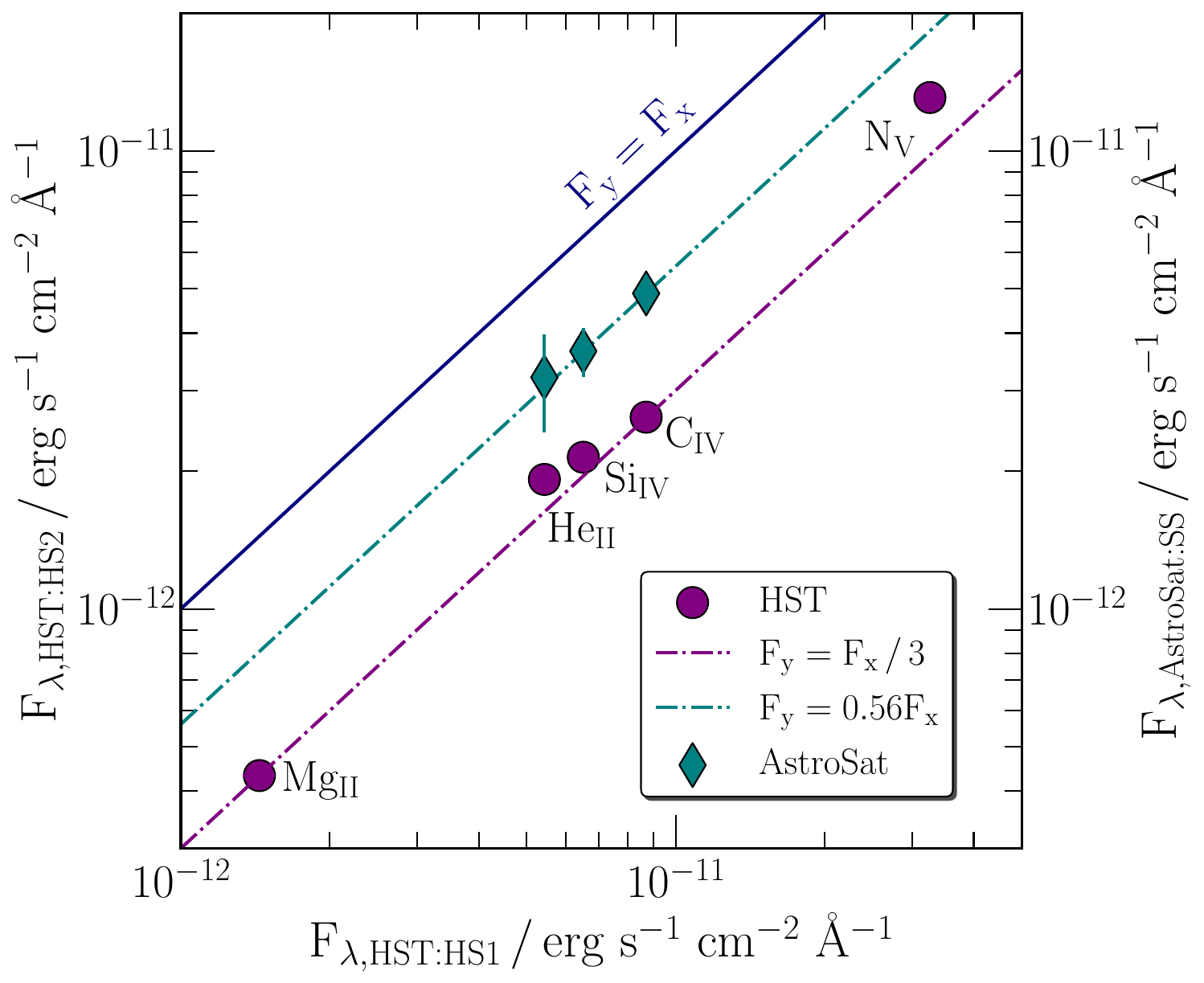}
\caption{The plot compares the line fluxes of \source~across our considered epochs. We use the HST:HS1 fluxes as our reference point along the x-axis and compare them with their HST:HS2 (purple colour) and AstroSat:SS (teal colour) counterparts. For a fair comparison between the \hst~and \astrosat~fluxes (as different settings are employed), we applied a correction to the AstroSat:SS fluxes. The correction includes the introduction of a factor, $\rm f_{corr}$, which is defined as the ratio of the line fluxes obtained from HST:HS1 to their downgraded counterparts. It is noted that the straight lines illustrate the relation between the different properties.}
\label{fig:MAXI1820_line_flux_evolution}
\end{figure}

Figure \ref{fig:MAXI1820_line_flux_evolution} compares the fluxes of the key lines between our three independent epochs: the luminous hard state (HST:HS1), the lower luminosity hard state just before the state transition (HST:HS2) and the soft state (AstroSat:SS). Two key features are apparent. First, the line fluxes decline by approximately a factor of 3 between the first and the second hard-state observations. This is similar to the drop in the UV continuum flux between these epochs, i.e. the EWs of the lines remain almost unchanged. Second, line fluxes then increase again -- by roughly a factor of 2 -- between HST:HS2 and the soft state. 

The strengthening of the UV lines across the hard-to-soft state transition may be expected, given the increase in EUV and soft X-ray luminosity across this transition (Figure \ref{fig:X-ray_lc_HID}). The line strength evolution between the two hard-state observations seems less obvious. Although the soft X-ray luminosity in HST:HS2 \textit{is} lower than that in HST:HS1 (Figure \ref{fig:X-ray_lc_HID}), the difference amounts to less than a factor of 2. Moreover, one might expect the strong hard X-ray component in HST:HS1 to \textit{inhibit} the formation of the UV lines, by overionizing the material in the disc atmosphere. 

However, it is important to remember that the radial locations of the line-forming regions are not the same in HST:HS1 and HST:HS2. Figure \ref{fig:MAXI1820_p2p_veloc_radii_potential} shows that all of the characteristic line-formation radii move \textit{inward} -- by slightly less than a factor of 2 -- between HST:HS1 and HST:HS2. This is consistent with each line being formed preferentially at a characteristic ionization parameter, $\rm U \propto L / n_e R^2$. If the density in the line-forming layers of the disc atmosphere is roughly constant, we expect the characteristic line-forming radius for a given line to scale with luminosity as $\rm R \propto L^{1/2}$. The evolution we see in line flux and velocity evolution between HST:HS1 and HST:HS2 is therefore likely due to the combination of these factors. 

\subsubsection{Absence of evidence for UV winds} \label{sec:discussion_UV_winds}

The classic observational wind signatures are blue-shifted absorption or P-Cygni line profiles, although these features are strongly inclination-dependent \citep[e.g.][]{ponti12, diaz-trigo14, diaz-trigo16}. The response of these signatures to X-ray luminosity variations and state transitions is of great interest, but difficult to explore. 

In this study, only our first hard state \hst~observation has the signal-to-noise and resolution to allow a sensitive search for UV wind signatures in \source. However, we find no evidence for these in the usual UV resonance lines (see Figure \ref{fig:MAXI_J1820_line_fits} for more information). 

The absence of outflow signatures prompts the question: is the wind truly absent or are we merely unable to detect it? Notably, winds in the hard state have been detected both in optical \citep{munoz-darias19} and near-infrared \citep{sanchez-sierras20} spectra of this source. These detections mainly manifested as P-Cygni and blue emission line wing features in Balmer/Helium and Paschen lines, respectively. 

Among known LMXBs, clear UV wind signatures have so far only been seen in Swift J1858.6-0814 \citep{castro-segura22} and UW CrB \citep{fijma23}. However, there are very few systems with UV observations in which such signatures could have been found. Moreover, the signatures in Swift J1858.6-0814 are weak and transient, making them quite difficult to detect. The absence of evidence for a "warm" outflow can therefore not yet be interpreted as evidence for its absence. More UV data $-$ for multiple systems and across different accretion states $-$ will be needed to address this issue.

\subsection{The evolutionary history of the binary} \label{sec:discussion_evolutionary_history}

The relative strengths of the UV resonance lines can serve as sensitive indicators of the physical conditions in the line-emitting gas. However, they can also be used as a complementary tool to determine the evolution of a system by estimating the initial mass of the donor and its evolutionary history. More specifically, the abundance and/or depletion of elements, such as \ion{N}{V} and \ion{C}{IV}, have been linked to the life stage and status of the companion. The key physics here is that the CNO cycle becomes the dominant nuclear process in stars with an initial mass of $\rm M_{2} \geq 1.4\Msun$ \citep{clayton83}. Hence the abundance $\rm \ion{N}{V}/\ion{C}{IV} $ ratio emerges as a reliable signature to discern whether the gas stream of the accreting material has undergone CNO-processing \citep[e.g.][]{mauche97, haswell02, gansicke03, froning11, froning14, castro-segura24}.

\begin{figure*}
\centering
\includegraphics[width=\textwidth]{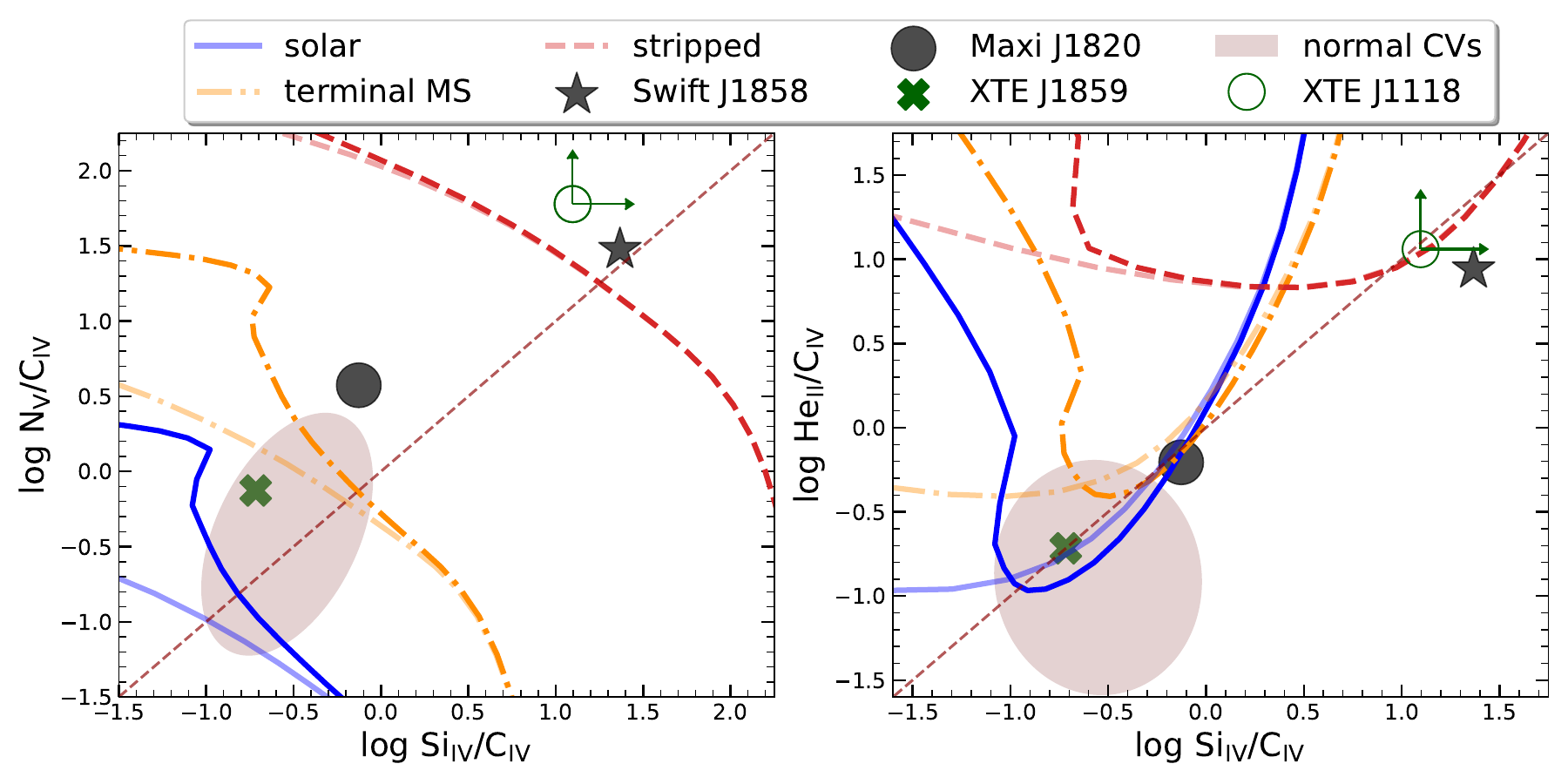}
\caption{Far-UV emission-line flux ratios for LMXBs: XTE J1118 + 480, XTE J1859 + 226 \citep{haswell02}, Swift J1858.6-0814 \citep{castro-segura24}. The region of the parameter space occupied by the `normal' CVs presented in \citet{mauche97} and \citet{gansicke03} is indicated with the shaded region, which encloses $2.5 \sigma$ of a two-dimensional (2D) Gaussian distribution. The lines are predicted line ratios as a function of the ionization parameter computed with {\sc cloudy} for an optically thin parcel of gas irradiated with a simple accretion disc presented above in Section \ref{sec:irradiated_model}. Solid, dash-dot and dashed lines are models carried out with solar abundances, $\rm M\approx1.5-2\Msun$ terminal main sequence (MS) and the equilibrium CNO-cycle core of a $\rm M \approx 2\Msun$ star, respectively. The latter is labelled as stripped, as we consider to have representative abundances of a stripped star with its convective CNO core exposed. The models include the \ion{O}{iv} multiplet within range of the \ion{Si}{iv} doublet. For reference, lines including the emission only from \ion{Si}{iv} are shown with the same linestyle but higher transparency. The measurements of \source~(filled black circle) at different epochs overlay in the plot but for clarity, a single measurement is shown. \source~lies very close to the theoretical predictions for a terminal MS star.}
\label{fig:MAXI1820_line_ratios}
\end{figure*}

The relevant line ratios of \source~are plotted in Figure \ref{fig:MAXI1820_line_ratios} along with the corresponding ratios for other systems, ranging from CVs to LMXBs. It appears that our source displays line ratios that are characteristic of "normal" CVs, suggesting that the accreting material has  \textit{not} undergone CNO processing. For reference, Figure \ref{fig:MAXI1820_line_ratios} also includes comparison systems with anomalous line ratios. In these systems, the donor star is usually thought to have had an initial mass $\gtrsim 2\Msun$. Its envelope was then stripped during a thermal timescale mass-transfer phase, leading to present-day surface abundances that reflect the earlier phase of CNO processing in the core \citep{schenker02}.

Given the well-determined distance ($\rm d=2.96 \pm 0.33$ kpc) of \source~\citep{atri20} and its low reddening value, there have already been efforts to determine its binary parameters. These suggest a BH primary with mass $\rm M_{BH} = (5.95 \pm 0.22)sin^{-3}i\ M_{\odot}$ and a K-type (subgiant) companion \citep{torres20,mikolajewska22}. Our results here imply that this donor is not the descendant of an initially much more massive star. 

\subsection{Is irradiation important?}
\label{sec:discussion_irradiation}

As shown in Section \ref{sec: Irradiated_disc_modelling}, the UV continuum shape, shown in Figure \ref{fig:MAXI1820_irradiation_modelling}, cannot be adequately described by a simple irradiated disc model with physically plausible parameters. This is mainly because the spectral shape is close to that of a viscously-dominated, unirradiated disc, similar to what happens in Nova Muscae 1991 \citep{cheng92} and A 0620-00 \citep{hynes05}. For physically reasonable parameters, the model is a poor match to the data, especially at the shortest far-UV wavelengths. Yet, X-ray irradiation is certainly \textit{expected} to heat up the outer parts of the disc. 

A related puzzle is that one might expect reprocessing to be quite sensitive to both the luminosity and SED of the X-ray radiation field. Yet the observed UV spectra are actually fairly similar across all three spectral states. The UV continuum luminosity tracks the drop in X-ray luminosity between the two hard states, but then actually stays constant across the hard-to-soft state transition. The emission line strengths also track the declining X-ray luminosity between the two hard states, before partially recovering in the soft state. However, the headline result is that both the continuum spectral shape and the dominant emission lines remain virtually unchanged, even as the system moves through three very different accretion states.

We can gain some insight into the implications of these observational findings by examining the physics underlying our simple irradiated disc model. In this model, X-rays are produced by point-like, isotropically emitting source at the center of the disc. Some of these X-ray photons impact on the disc surface and are absorbed there. This heats up the irradiated parts of the disc, thus increasing the effective temperature and modifying the radial intensity profile. 

In reality, the details of this process are complex \citep[e.g.][]{dubus99}. The geometry and location of the X-ray emitting "corona" are highly uncertain, but clearly important. Similarly, the shape and scale height of the disc -- $\rm H/R|_R$ -- are key factors in determining what fraction of X-ray photons are intercepted by the disc. However, the relevant scale height here is \textit{not} really the pressure scale height, but rather the height above the disc at which it presents an optical depth of $\simeq 1$ \textit{to the irradiating photons}. Finally, the detailed physics governing the reprocessing of X-rays in the disc atmosphere are clearly far more complex than what can be captured by a single efficiency parameter (i.e. the albedo). 

It is easy to see that some or all of these processes are expected to depend on the accretion state of the source. For example, the opacity the disc atmosphere presents to incoming X-ray photons, as well as the efficiency with which those photons are reprocessed, are sensitive to whether these photons are soft or hard. And yet the presumably reprocessing-dominated UV emission-line spectrum remains qualitatively unchanged. 

Part of the explanation may be that the relevant parts of the disc do not see the same X-ray SED that we do, and that the formation of UV emission lines is mainly controlled by EUV photons. For reference, by "EUV", here we mean photons with energy between, say, 13.6~eV and 100~eV (check Figure \ref{fig:MAXI1820_p2p_veloc_radii_potential}, which includes the ionization potentials of the relevant species). In fact, the \ion{He}{ii} recombination line is generally thought to be a reliable bolometer for the EUV luminosity above 54~eV (or, in this case, the fraction of this luminosity that is intercepted by the disc). 

As noted above, the \ion{He}{ii} emission line flux drops by $\simeq \times 3$ between the luminous and less luminous hard states (HST:HS1 and HST:HS2, respectively) before partially recovering in the soft state (AstroSat:SS). On the one hand, these relatively modest variations  are in line with the modest changes in the observed emission line spectra. And indeed the soft X-ray flux also only varies by a comparable amount (see Figure \ref{fig:X-ray_lc_HID}). 

On the other hand, the \ion{He}{ii} line flux is clearly not proportional to the soft X-ray flux: the latter is comparable in the two hard states, but a factor of $\simeq \times 3$ higher in the soft state. Moreover, the \textit{hard} X-ray component completely dominates in HST:HS1, is comparable to the soft X-ray component in HST:HS2, and is all but completely quenched in AstroSat:SS. So, somehow, the UV line-forming regions must be shielded from the ionizing effects of the energetic X-ray photons produced in the hard state. 

Our main conclusion from all these considerations is that the physics governing the reprocessing of X-rays in the outer disc regions is (a) complex and (b) poorly described by simple models. We therefore strongly encourage theoretical efforts to construct physically realistic models of irradiated accretion disc atmospheres.

\subsection{Evidence of a UV QPO signal?}  \label{sec:discussion_QPOs}

Only a handful of systems to date are known to display LF-QPOs in multiple wave bands \citep[mostly in the optical/infrared and in X-rays; check][for more information]{motch82, motch83, hynes03, durant09, gandhi10}. The only system known to exhibit an analogous signal in the UV is the BHXT XTE J1118+480 \citep{hynes03}. The mentioned studies are based on panchromatic, (quasi-)simultaneous observations of BHXTs and correspond to periods when the systems are in the low hard state. It is still unclear, though, whether the X-ray and longer-wavelength signals are physically associated and how the latter are actually produced \citep[e.g.][]{markoff01,  hynes03, veledina13, veledina15, gandhi17}.

In this work, our time-resolved observations allow us to search for UV QPO signatures only in the \textit{hard} state. Nevertheless, \source~is already known to display simultaneous X-ray and optical LF-QPOs in the hard state, as reported by \citet{mao22} and \citet{thomas22}. In this state, the observed X-ray QPOs, the so-called type-C QPOs, may be associated with a precessing hot flow near the inner edge of the disc \citep[e.g.][]{stella98, ingram09, ingram19}, instabilities in the disc \citep[e.g.][]{tagger99, varniere02, varniere12} or corona variability \citep[e.g.][]{titarchuk04, cabanac10}. On the other hand, in the optical band, detections of QPOs are limited, and the observed signals are usually attributed to precession \citep{veledina13}, thermal reprocessing \citep{veledina15} or jet synchrotron emission variations  \citep[e.g.][]{markoff01, hynes03, gandhi17}.

Almost at the time of our observations, \citet{mao22} and \citet{thomas22} find that both X-ray and optical signals share similar centroid QPO frequencies and evolve similarly over time. However, optical variations are more coherent, i.e. $\rm Q_{X-ray} < Q_{opt}$ \citep[check Table 1 in][]{mao22}. This implies that disc reprocessing is not the dominant mechanism, and \citet{ma21} suggest instead that the jet may precess at the QPO frequency. 

Our work provides provisional evidence for the existence of UV LF-QPOs with centroid frequencies closely resembling those observed in other bands. Our modelling suggests $-$ albeit  taking into account only the observations when we \textit{do} detect a QPO $-$ that the coherence in the UV is lower than the equivalent one in optical and X-rays (for almost the same centroid frequencies). It seems unlikely that synchrotron radiation could be responsible for QPOs in the UV band. The current results suggest that the "same" type-C QPOs in different bands can be produced in three distinct regions $-$ the inner accretion flow (producing X-ray QPOs), the jet (producing optical QPOs) and the outer disc (producing UV QPOs via reprocessing). In the UV, reprocessing may constitute the primary mechanism, which is in line with the lower coherence we observe. 

\section{Summary}

We have presented the first multi-epoch, time-resolved and spectrally-resolved UV characterization of the transient X-ray binary \source. We obtained observations in three distinct stages of the outburst: a luminous hard state after the eruption peak, a less luminous hard state just before the state transition and finally the soft state.

Our main conclusions are the following:
\begin{itemize}
    \item We have determined the interstellar reddening and extinction towards the source via the $\lambda2175$\AA~absorption feature (near-UV bump) and Ly$\alpha$ modelling. Our estimate of the reddening is quoted as $\rm E_{B-V} = 0.20\pm 0.05$. 
    \item We track the spectral evolution of \source~throughout its outburst. Surprisingly, we see no major differences in the appearance of the UV emission line spectrum across the three distinct spectral states. The UV spectra are characterized by blue continua with superposed broad, double-peaked emission lines,  such as \ion{N}{v} $\lambda1240$\AA, \ion{Si}{iv} $\lambda1400$\AA, \ion{C}{iv} $\lambda1550$\AA, \ion{He}{ii} $\lambda1640$\AA, \ion{Mg}{ii} $\lambda2800$\AA. 
    \item We do not find evidence of an outflow in the form of blue-shifted absorption or P-Cygni profiles in any of the considered lines that are mentioned above. 
    \item We use the relative strengths of the UV resonance lines to constrain the evolutionary history of the binary, showing that the donor had not undergone CNO processing in the past and presumably now lies at the end of the MS or at the subgiant branch. 
    \item Simple irradiated disc models with physically plausible parameters fail to adequately describe the observed UV continuum shape. This is mainly because the observed spectral slope is close to that for an unirradiated, viscously-dominated accretion disc.
    \item There is marginal evidence for a UV LF-QPO signature in the data obtained during the luminous hard state (HST:HS1). The frequencies of these candidate QPOs are comparable to those of QPOs detected almost at the same time in X-ray and optical observations of \source. This provisional evidence is worthy of consideration and if confirmed, these signatures may be produced through reprocessing at the outer disc regions. 
\end{itemize}

\section*{Acknowledgements}

We would like to thank Dr.Douglas Buisson for providing us with the \nicer~data products and Dr.Andrés Gúrpide Lasheras for helpful discussions. 
MG acknowledges support by the Mayflower scholarship in the School of Physics and Astronomy of the University of Southampton. Partial support for KSL's effort on the project was provided by NASA through grant numbers HST-GO-16489 and HST-GO-16659 from the Space Telescope Science Institute, which is operated by AURA, Inc., under NASA contract NAS 5-26555. RIH acknowledges support from NASA through grant number HST-GO-15454 from the Space Telescope Science Institute, which is operated by AURA, Inc., under NASA contract NAS 5-26555. 

This work was conducted with the help of {\sc numpy} \citep{vanderwalt11}, {\sc astropy} \citep{astropy18}, {\sc scipy} \citep{scipy2020} and {\sc matplotlib} \citep{hunter07}. 

\section*{Data Availability}

The main datasets underlying this study are publicly available through their respective archives: 

\hst/MAST archive $-$ {\small \url{https://archive.stsci.edu/}}

\astrosat/ISSDC archive $-$ {\small \url{https://astrobrowse.issdc.gov.in/astro_sarchive/archive/Home.jsp}}

Complimentary datasets in X-rays (\swift/\nicer) and optical (\textit{VLT}) are found at the HEASARC ({\small \url{heasarc.gsfc.nasa.gov}}) and ESO webpages ({\small \url{http://archive.eso.org}}), respectively.



\bibliographystyle{mnras}
\bibliography{biblio} 

\begin{thebibliography}{}
\makeatletter
\relax
\def\mn@urlcharsother{\let\do\@makeother \do\$\do\&\do\#\do\^\do\_\do\%\do\~}
\def\mn@doi{\begingroup\mn@urlcharsother \@ifnextchar [ {\mn@doi@} {\mn@doi@[]}}
\def\mn@doi@[#1]#2{\def\@tempa{#1}\ifx\@tempa\@empty \href {http://dx.doi.org/#2} {doi:#2}\else \href {http://dx.doi.org/#2} {#1}\fi \endgroup}
\def\mn@eprint#1#2{\mn@eprint@#1:#2::\@nil}
\def\mn@eprint@arXiv#1{\href {http://arxiv.org/abs/#1} {{\tt arXiv:#1}}}
\def\mn@eprint@dblp#1{\href {http://dblp.uni-trier.de/rec/bibtex/#1.xml} {dblp:#1}}
\def\mn@eprint@#1:#2:#3:#4\@nil{\def\@tempa {#1}\def\@tempb {#2}\def\@tempc {#3}\ifx \@tempc \@empty \let \@tempc \@tempb \let \@tempb \@tempa \fi \ifx \@tempb \@empty \def\@tempb {arXiv}\fi \@ifundefined {mn@eprint@\@tempb}{\@tempb:\@tempc}{\expandafter \expandafter \csname mn@eprint@\@tempb\endcsname \expandafter{\@tempc}}}

\bibitem[\protect\citeauthoryear{{Astropy Collaboration} et~al.,}{{Astropy Collaboration} et~al.}{2018}]{astropy18}
{Astropy Collaboration} et~al., 2018, \mn@doi [\aj] {10.3847/1538-3881/aabc4f}, \href {https://ui.adsabs.harvard.edu/abs/2018AJ....156..123A} {156, 123}

\bibitem[\protect\citeauthoryear{{Atri} et~al.,}{{Atri} et~al.}{2020}]{atri20}
{Atri} P.,  et~al., 2020, \mn@doi [\mnras] {10.1093/mnrasl/slaa010}, \href {https://ui.adsabs.harvard.edu/abs/2020MNRAS.493L..81A} {493, L81}

\bibitem[\protect\citeauthoryear{{Baglio}, {Russell}  \& {Lewis}}{{Baglio} et~al.}{2018}]{baglio18}
{Baglio} M.~C.,  {Russell} D.~M.,   {Lewis} F.,  2018, The Astronomer's Telegram, \href {https://ui.adsabs.harvard.edu/abs/2018ATel11418....1B} {11418, 1}

\bibitem[\protect\citeauthoryear{Banerjee et~al.,}{Banerjee et~al.}{2024}]{banerjee24}
Banerjee S.,  et~al., 2024, \mn@doi [The Astrophysical Journal] {10.3847/1538-4357/ad24ef}, 964, 189

\bibitem[\protect\citeauthoryear{{Belloni}}{{Belloni}}{2010}]{belloni10}
{Belloni} T.~M.,  2010, in {Belloni} T.,  ed., Lecture Notes in Physics, Vol.~794, The Jet Paradigm: From Microquasars to Quasars.
Springer Berlin Heidelberg, p.~53

\bibitem[\protect\citeauthoryear{{Belloni} \& {Motta}}{{Belloni} \& {Motta}}{2016}]{belloni16}
{Belloni} T.~M.,  {Motta} S.~E.,  2016, in {Bambi} C.,  ed.,  Astrophysics and Space Science Library Vol. 440, Astrophysics of Black Holes: From Fundamental Aspects to Latest Developments. p.~61

\bibitem[\protect\citeauthoryear{{Belloni} \& {Stella}}{{Belloni} \& {Stella}}{2014}]{belloni14}
{Belloni} T.~M.,  {Stella} L.,  2014, \mn@doi [\ssr] {10.1007/s11214-014-0076-0}, \href {https://ui.adsabs.harvard.edu/abs/2014SSRv..183...43B} {183, 43}

\bibitem[\protect\citeauthoryear{{Belloni}, {Homan}, {Casella}, {van der Klis}, {Nespoli}, {Lewin}, {Miller}  \& {M{\'e}ndez}}{{Belloni} et~al.}{2005}]{belloni05}
{Belloni} T.,  {Homan} J.,  {Casella} P.,  {van der Klis} M.,  {Nespoli} E.,  {Lewin} W.~H.~G.,  {Miller} J.~M.,   {M{\'e}ndez} M.,  2005, \mn@doi [\aap] {10.1051/0004-6361:20042457}, \href {https://ui.adsabs.harvard.edu/abs/2005A&A...440..207B} {440, 207}

\bibitem[\protect\citeauthoryear{{Bohlin}}{{Bohlin}}{1975}]{bohlin75}
{Bohlin} R.~C.,  1975, \mn@doi [\apj] {10.1086/153803}, \href {https://ui.adsabs.harvard.edu/abs/1975ApJ...200..402B} {200, 402}

\bibitem[\protect\citeauthoryear{{Bohlin}, {Savage}  \& {Drake}}{{Bohlin} et~al.}{1978}]{bohlin78}
{Bohlin} R.~C.,  {Savage} B.~D.,   {Drake} J.~F.,  1978, \mn@doi [\apj] {10.1086/156357}, \href {https://ui.adsabs.harvard.edu/abs/1978ApJ...224..132B} {224, 132}

\bibitem[\protect\citeauthoryear{{Cabanac}, {Henri}, {Petrucci}, {Malzac}, {Ferreira}  \& {Belloni}}{{Cabanac} et~al.}{2010}]{cabanac10}
{Cabanac} C.,  {Henri} G.,  {Petrucci} P.~O.,  {Malzac} J.,  {Ferreira} J.,   {Belloni} T.~M.,  2010, \mn@doi [\mnras] {10.1111/j.1365-2966.2010.16340.x}, \href {https://ui.adsabs.harvard.edu/abs/2010MNRAS.404..738C} {404, 738}

\bibitem[\protect\citeauthoryear{{Casella}, {Belloni}, {Homan}  \& {Stella}}{{Casella} et~al.}{2004}]{casella04}
{Casella} P.,  {Belloni} T.,  {Homan} J.,   {Stella} L.,  2004, \mn@doi [\aap] {10.1051/0004-6361:20041231}, \href {https://ui.adsabs.harvard.edu/abs/2004A&A...426..587C} {426, 587}

\bibitem[\protect\citeauthoryear{{Casella}, {Testa}, {Russell}, {Belloni}  \& {Maccarone}}{{Casella} et~al.}{2018}]{casella18}
{Casella} P.,  {Testa} V.,  {Russell} D.~M.,  {Belloni} T.~M.,   {Maccarone} T.~J.,  2018, The Astronomer's Telegram, \href {https://ui.adsabs.harvard.edu/abs/2018ATel11833....1C} {11833, 1}

\bibitem[\protect\citeauthoryear{{Castro Segura} et~al.,}{{Castro Segura} et~al.}{2022}]{castro-segura22}
{Castro Segura} N.,  et~al., 2022, \mn@doi [\nat] {10.1038/s41586-021-04324-2}, \href {https://ui.adsabs.harvard.edu/abs/2022Natur.603...52C} {603, 52}

\bibitem[\protect\citeauthoryear{{Castro Segura} et~al.,}{{Castro Segura} et~al.}{2024}]{castro-segura24}
{Castro Segura} N.,  et~al., 2024, \mn@doi [\mnras] {10.1093/mnras/stad3109}, \href {https://ui.adsabs.harvard.edu/abs/2024MNRAS.527.2508C} {527, 2508}

\bibitem[\protect\citeauthoryear{{Cheng}, {Horne}, {Panagia}, {Shrader}, {Gilmozzi}, {Paresce}  \& {Lund}}{{Cheng} et~al.}{1992}]{cheng92}
{Cheng} F.~H.,  {Horne} K.,  {Panagia} N.,  {Shrader} C.~R.,  {Gilmozzi} R.,  {Paresce} F.,   {Lund} N.,  1992, \mn@doi [\apj] {10.1086/171822}, \href {https://ui.adsabs.harvard.edu/abs/1992ApJ...397..664C} {397, 664}

\bibitem[\protect\citeauthoryear{Clayton}{Clayton}{1983}]{clayton83}
Clayton D.~D.,  1983, Principles of stellar evolution and nucleosynthesis.
University of Chicago press

\bibitem[\protect\citeauthoryear{{Corbel} et~al.,}{{Corbel} et~al.}{2001}]{corbel01}
{Corbel} S.,  et~al., 2001, \mn@doi [\apj] {10.1086/321364}, \href {https://ui.adsabs.harvard.edu/abs/2001ApJ...554...43C} {554, 43}

\bibitem[\protect\citeauthoryear{{Corbel}, {Nowak}, {Fender}, {Tzioumis}  \& {Markoff}}{{Corbel} et~al.}{2003}]{corbel03}
{Corbel} S.,  {Nowak} M.~A.,  {Fender} R.~P.,  {Tzioumis} A.~K.,   {Markoff} S.,  2003, \mn@doi [\aap] {10.1051/0004-6361:20030090}, \href {https://ui.adsabs.harvard.edu/abs/2003A&A...400.1007C} {400, 1007}

\bibitem[\protect\citeauthoryear{{Corral-Santana}, {Casares}, {Mu{\~n}oz-Darias}, {Bauer}, {Mart{\'\i}nez-Pais}  \& {Russell}}{{Corral-Santana} et~al.}{2016}]{corral-santana16}
{Corral-Santana} J.~M.,  {Casares} J.,  {Mu{\~n}oz-Darias} T.,  {Bauer} F.~E.,  {Mart{\'\i}nez-Pais} I.~G.,   {Russell} D.~M.,  2016, \mn@doi [\aap] {10.1051/0004-6361/201527130}, \href {https://ui.adsabs.harvard.edu/abs/2016A&A...587A..61C} {587, A61}

\bibitem[\protect\citeauthoryear{{Davis} \& {El-Abd}}{{Davis} \& {El-Abd}}{2019}]{davis19}
{Davis} S.~W.,  {El-Abd} S.,  2019, \mn@doi [\apj] {10.3847/1538-4357/ab05c5}, \href {https://ui.adsabs.harvard.edu/abs/2019ApJ...874...23D} {874, 23}

\bibitem[\protect\citeauthoryear{{Dewangan}}{{Dewangan}}{2021}]{dewangan2021}
{Dewangan} G.~C.,  2021, \mn@doi [Journal of Astrophysics and Astronomy] {10.1007/s12036-021-09691-w}, \href {https://ui.adsabs.harvard.edu/abs/2021JApA...42...49D} {42, 49}

\bibitem[\protect\citeauthoryear{{D{\'\i}az Trigo} \& {Boirin}}{{D{\'\i}az Trigo} \& {Boirin}}{2016}]{diaz-trigo16}
{D{\'\i}az Trigo} M.,  {Boirin} L.,  2016, \mn@doi [Astronomische Nachrichten] {10.1002/asna.201612315}, \href {https://ui.adsabs.harvard.edu/abs/2016AN....337..368D} {337, 368}

\bibitem[\protect\citeauthoryear{{D{\'\i}az Trigo}, {Migliari}, {Miller-Jones}  \& {Guainazzi}}{{D{\'\i}az Trigo} et~al.}{2014}]{diaz-trigo14}
{D{\'\i}az Trigo} M.,  {Migliari} S.,  {Miller-Jones} J.~C.~A.,   {Guainazzi} M.,  2014, \mn@doi [\aap] {10.1051/0004-6361/201424554}, \href {https://ui.adsabs.harvard.edu/abs/2014A&A...571A..76D} {571, A76}

\bibitem[\protect\citeauthoryear{{Done}, {Gierli{\'n}ski}  \& {Kubota}}{{Done} et~al.}{2007}]{done07}
{Done} C.,  {Gierli{\'n}ski} M.,   {Kubota} A.,  2007, \mn@doi [\aapr] {10.1007/s00159-007-0006-1}, \href {https://ui.adsabs.harvard.edu/abs/2007A&ARv..15....1D} {15, 1}

\bibitem[\protect\citeauthoryear{{Dubus}, {Lasota}, {Hameury}  \& {Charles}}{{Dubus} et~al.}{1999}]{dubus99}
{Dubus} G.,  {Lasota} J.-P.,  {Hameury} J.-M.,   {Charles} P.,  1999, \mn@doi [\mnras] {10.1046/j.1365-8711.1999.02212.x}, \href {https://ui.adsabs.harvard.edu/abs/1999MNRAS.303..139D} {303, 139}

\bibitem[\protect\citeauthoryear{{Dubus}, {Hameury}  \& {Lasota}}{{Dubus} et~al.}{2001}]{dubus01}
{Dubus} G.,  {Hameury} J.~M.,   {Lasota} J.~P.,  2001, \mn@doi [\aap] {10.1051/0004-6361:20010632}, \href {https://ui.adsabs.harvard.edu/abs/2001A&A...373..251D} {373, 251}

\bibitem[\protect\citeauthoryear{{Durant}, {Gandhi}, {Shahbaz}, {Peralta}  \& {Dhillon}}{{Durant} et~al.}{2009}]{durant09}
{Durant} M.,  {Gandhi} P.,  {Shahbaz} T.,  {Peralta} H.~H.,   {Dhillon} V.~S.,  2009, \mn@doi [\mnras] {10.1111/j.1365-2966.2008.14044.x}, \href {https://ui.adsabs.harvard.edu/abs/2009MNRAS.392..309D} {392, 309}

\bibitem[\protect\citeauthoryear{Esin, McClintock  \& Narayan}{Esin et~al.}{1997}]{esin97}
Esin A.~A.,  McClintock J.~E.,   Narayan R.,  1997, \mn@doi [\apj] {10.1086/304829}, 489, 865

\bibitem[\protect\citeauthoryear{{Fender} \& {Gallo}}{{Fender} \& {Gallo}}{2014}]{fender14}
{Fender} R.,  {Gallo} E.,  2014, \mn@doi [\ssr] {10.1007/s11214-014-0069-z}, \href {https://ui.adsabs.harvard.edu/abs/2014SSRv..183..323F} {183, 323}

\bibitem[\protect\citeauthoryear{{Fender} et~al.,}{{Fender} et~al.}{1999}]{fender99}
{Fender} R.,  et~al., 1999, \mn@doi [\apjl] {10.1086/312128}, \href {https://ui.adsabs.harvard.edu/abs/1999ApJ...519L.165F} {519, L165}

\bibitem[\protect\citeauthoryear{{Fender}, {Belloni}  \& {Gallo}}{{Fender} et~al.}{2004}]{fender04}
{Fender} R.~P.,  {Belloni} T.~M.,   {Gallo} E.,  2004, \mn@doi [\mnras] {10.1111/j.1365-2966.2004.08384.x}, \href {https://ui.adsabs.harvard.edu/abs/2004MNRAS.355.1105F} {355, 1105}

\bibitem[\protect\citeauthoryear{{Fender}, {Homan}  \& {Belloni}}{{Fender} et~al.}{2009}]{fender09}
{Fender} R.~P.,  {Homan} J.,   {Belloni} T.~M.,  2009, \mn@doi [\mnras] {10.1111/j.1365-2966.2009.14841.x}, \href {https://ui.adsabs.harvard.edu/abs/2009MNRAS.396.1370F} {396, 1370}

\bibitem[\protect\citeauthoryear{{Ferreira} et~al.,}{{Ferreira} et~al.}{2022}]{ferreira22}
{Ferreira} J.,  et~al., 2022, \mn@doi [\aap] {10.1051/0004-6361/202040165}, \href {https://ui.adsabs.harvard.edu/abs/2022A&A...660A..66F} {660, A66}

\bibitem[\protect\citeauthoryear{{Fijma}, {Castro Segura}, {Degenaar}, {Knigge}, {Higginbottom}, {Hern{\'a}ndez Santisteban}  \& {Maccarone}}{{Fijma} et~al.}{2023}]{fijma23}
{Fijma} S.,  {Castro Segura} N.,  {Degenaar} N.,  {Knigge} C.,  {Higginbottom} N.,  {Hern{\'a}ndez Santisteban} J.~V.,   {Maccarone} T.~J.,  2023, \mn@doi [\mnras] {10.1093/mnrasl/slad125}, \href {https://ui.adsabs.harvard.edu/abs/2023MNRAS.526L.149F} {526, L149}

\bibitem[\protect\citeauthoryear{{Fitzpatrick}}{{Fitzpatrick}}{1999}]{fitzpatrick99}
{Fitzpatrick} E.~L.,  1999, \mn@doi [\pasp] {10.1086/316293}, \href {https://ui.adsabs.harvard.edu/abs/1999PASP..111...63F} {111, 63}

\bibitem[\protect\citeauthoryear{Frank, King  \& Raine}{Frank et~al.}{2002}]{frank02}
Frank J.,  King A.,   Raine D.,  2002, Accretion Power in Astrophysics: Third edition.
Cambridge University Press, New York

\bibitem[\protect\citeauthoryear{{Froning} et~al.,}{{Froning} et~al.}{2011}]{froning11}
{Froning} C.~S.,  et~al., 2011, \mn@doi [\apj] {10.1088/0004-637X/743/1/26}, \href {https://ui.adsabs.harvard.edu/abs/2011ApJ...743...26F} {743, 26}

\bibitem[\protect\citeauthoryear{{Froning}, {Maccarone}, {France}, {Winter}, {Robinson}, {Hynes}  \& {Lewis}}{{Froning} et~al.}{2014}]{froning14}
{Froning} C.~S.,  {Maccarone} T.~J.,  {France} K.,  {Winter} L.,  {Robinson} E.~L.,  {Hynes} R.~I.,   {Lewis} F.,  2014, \mn@doi [\apj] {10.1088/0004-637X/780/1/48}, \href {https://ui.adsabs.harvard.edu/abs/2014ApJ...780...48F} {780, 48}

\bibitem[\protect\citeauthoryear{{Gandhi} et~al.,}{{Gandhi} et~al.}{2010}]{gandhi10}
{Gandhi} P.,  et~al., 2010, \mn@doi [\mnras] {10.1111/j.1365-2966.2010.17083.x}, \href {https://ui.adsabs.harvard.edu/abs/2010MNRAS.407.2166G} {407, 2166}

\bibitem[\protect\citeauthoryear{{Gandhi} et~al.,}{{Gandhi} et~al.}{2017}]{gandhi17}
{Gandhi} P.,  et~al., 2017, \mn@doi [Nature Astronomy] {10.1038/s41550-017-0273-3}, \href {https://ui.adsabs.harvard.edu/abs/2017NatAs...1..859G} {1, 859}

\bibitem[\protect\citeauthoryear{{G{\"a}nsicke} et~al.,}{{G{\"a}nsicke} et~al.}{2003}]{gansicke03}
{G{\"a}nsicke} B.~T.,  et~al., 2003, \apj, \href {https://ui.adsabs.harvard.edu/abs/2003ApJ...594..443G} {594, 443}

\bibitem[\protect\citeauthoryear{{Garcia}, {Brown}, {Pahre}, {McClintock}, {Callanan}  \& {Garnavich}}{{Garcia} et~al.}{2000}]{Garcia00}
{Garcia} M.,  {Brown} W.,  {Pahre} M.,  {McClintock} J.,  {Callanan} P.,   {Garnavich} P.,  2000, \iaucirc, \href {https://ui.adsabs.harvard.edu/abs/2000IAUC.7392....2G} {7392, 2}

\bibitem[\protect\citeauthoryear{{Gehrels} et~al.,}{{Gehrels} et~al.}{2004}]{gehrels04}
{Gehrels} N.,  et~al., 2004, \mn@doi [\apj] {10.1086/422091}, \href {https://ui.adsabs.harvard.edu/abs/2004ApJ...611.1005G} {611, 1005}

\bibitem[\protect\citeauthoryear{Gendreau et~al.,}{Gendreau et~al.}{2016}]{gendreau16}
Gendreau K.~C.,  et~al., 2016, \mn@doi [SPIE Proceedings] {10.1117/12.2231304}, 9905, 420

\bibitem[\protect\citeauthoryear{{Gilfanov}}{{Gilfanov}}{2010}]{gilfanov10}
{Gilfanov} M.,  2010, in {Belloni} T.,  ed., Lecture Notes in Physics, Vol.~794, The Jet Paradigm: From Microquasars to Quasars.
Springer Berlin Heidelberg, p.~17

\bibitem[\protect\citeauthoryear{{Grove}, {Johnson}, {Kroeger}, {McNaron-Brown}, {Skibo}  \& {Phlips}}{{Grove} et~al.}{1998}]{grove98}
{Grove} J.~E.,  {Johnson} W.~N.,  {Kroeger} R.~A.,  {McNaron-Brown} K.,  {Skibo} J.~G.,   {Phlips} B.~F.,  1998, \mn@doi [\apj] {10.1086/305746}, \href {https://ui.adsabs.harvard.edu/abs/1998ApJ...500..899G} {500, 899}

\bibitem[\protect\citeauthoryear{{Guan} et~al.,}{{Guan} et~al.}{2021}]{guan21}
{Guan} J.,  et~al., 2021, \mn@doi [\mnras] {10.1093/mnras/stab945}, \href {https://ui.adsabs.harvard.edu/abs/2021MNRAS.504.2168G} {504, 2168}

\bibitem[\protect\citeauthoryear{{HI4PI Collaboration} et~al.,}{{HI4PI Collaboration} et~al.}{2016}]{h14pi_collaboration16}
{HI4PI Collaboration} et~al., 2016, \mn@doi [\aap] {10.1051/0004-6361/201629178}, \href {https://ui.adsabs.harvard.edu/abs/2016A&A...594A.116H} {594, A116}

\bibitem[\protect\citeauthoryear{{Hameury}}{{Hameury}}{2020}]{hameury20}
{Hameury} J.~M.,  2020, Advances in Space Research, 66, 1004

\bibitem[\protect\citeauthoryear{{Haswell}, {Hynes}, {King}  \& {Schenker}}{{Haswell} et~al.}{2002}]{haswell02}
{Haswell} C.~A.,  {Hynes} R.~I.,  {King} A.~R.,   {Schenker} K.,  2002, \mn@doi [\mnras] {10.1046/j.1365-8711.2002.05369.x}, \href {https://ui.adsabs.harvard.edu/abs/2002MNRAS.332..928H} {332, 928}

\bibitem[\protect\citeauthoryear{{Higginbottom}, {Knigge}, {Long}, {Matthews}, {Sim}  \& {Hewitt}}{{Higginbottom} et~al.}{2018}]{higginbottom18}
{Higginbottom} N.,  {Knigge} C.,  {Long} K.~S.,  {Matthews} J.~H.,  {Sim} S.~A.,   {Hewitt} H.~A.,  2018, \mn@doi [\mnras] {10.1093/mnras/sty1599}, \href {https://ui.adsabs.harvard.edu/abs/2018MNRAS.479.3651H} {479, 3651}

\bibitem[\protect\citeauthoryear{{Higginbottom}, {Knigge}, {Long}, {Matthews}  \& {Parkinson}}{{Higginbottom} et~al.}{2019}]{higginbottom19}
{Higginbottom} N.,  {Knigge} C.,  {Long} K.~S.,  {Matthews} J.~H.,   {Parkinson} E.~J.,  2019, \mn@doi [\mnras] {10.1093/mnras/stz310}, \href {https://ui.adsabs.harvard.edu/abs/2019MNRAS.484.4635H} {484, 4635}

\bibitem[\protect\citeauthoryear{{Higginbottom}, {Knigge}, {Sim}, {Long}, {Matthews}, {Hewitt}, {Parkinson}  \& {Mangham}}{{Higginbottom} et~al.}{2020}]{higginbottom20}
{Higginbottom} N.,  {Knigge} C.,  {Sim} S.~A.,  {Long} K.~S.,  {Matthews} J.~H.,  {Hewitt} H.~A.,  {Parkinson} E.~J.,   {Mangham} S.~W.,  2020, \mn@doi [\mnras] {10.1093/mnras/staa209}, \href {https://ui.adsabs.harvard.edu/abs/2020MNRAS.492.5271H} {492, 5271}

\bibitem[\protect\citeauthoryear{{Homan} \& {Belloni}}{{Homan} \& {Belloni}}{2005}]{homan05}
{Homan} J.,  {Belloni} T.,  2005, Astrophysics and Space Science, \href {https://ui.adsabs.harvard.edu/abs/2005Ap&SS.300..107H} {300, 107}

\bibitem[\protect\citeauthoryear{{Homan}, {Wijnands}, {van der Klis}, {Belloni}, {van Paradijs}, {Klein-Wolt}, {Fender}  \& {M{\'e}ndez}}{{Homan} et~al.}{2001}]{homan01}
{Homan} J.,  {Wijnands} R.,  {van der Klis} M.,  {Belloni} T.,  {van Paradijs} J.,  {Klein-Wolt} M.,  {Fender} R.,   {M{\'e}ndez} M.,  2001, \mn@doi [\apjs] {10.1086/318954}, \href {https://ui.adsabs.harvard.edu/abs/2001ApJS..132..377H} {132, 377}

\bibitem[\protect\citeauthoryear{{Homan} et~al.,}{{Homan} et~al.}{2018a}]{homan18a}
{Homan} J.,  et~al., 2018a, The Astronomer's Telegram, \href {https://ui.adsabs.harvard.edu/abs/2018ATel11820....1H} {11820, 1}

\bibitem[\protect\citeauthoryear{{Homan} et~al.,}{{Homan} et~al.}{2018b}]{homan18b}
{Homan} J.,  et~al., 2018b, The Astronomer's Telegram, \href {https://ui.adsabs.harvard.edu/abs/2018ATel11823....1H} {11823, 1}

\bibitem[\protect\citeauthoryear{{Homan} et~al.,}{{Homan} et~al.}{2020}]{homan20}
{Homan} J.,  et~al., 2020, \mn@doi [\apjl] {10.3847/2041-8213/ab7932}, \href {https://ui.adsabs.harvard.edu/abs/2020ApJ...891L..29H} {891, L29}

\bibitem[\protect\citeauthoryear{{Hunter}}{{Hunter}}{2007}]{hunter07}
{Hunter} J.~D.,  2007, \mn@doi [Computing in Science and Engineering] {10.1109/MCSE.2007.55}, \href {https://ui.adsabs.harvard.edu/abs/2007CSE.....9...90H} {9, 90}

\bibitem[\protect\citeauthoryear{{Hynes}}{{Hynes}}{2005}]{hynes05}
{Hynes} R.~I.,  2005, \mn@doi [\apj] {10.1086/428445}, \href {https://ui.adsabs.harvard.edu/abs/2005ApJ...623.1026H} {623, 1026}

\bibitem[\protect\citeauthoryear{{Hynes}, {Haswell}, {Chaty}, {Shrader}  \& {Cui}}{{Hynes} et~al.}{2002}]{hynes02}
{Hynes} R.~I.,  {Haswell} C.~A.,  {Chaty} S.,  {Shrader} C.~R.,   {Cui} W.,  2002, \mn@doi [\mnras] {10.1046/j.1365-8711.2002.05175.x}, \href {https://ui.adsabs.harvard.edu/abs/2002MNRAS.331..169H} {331, 169}

\bibitem[\protect\citeauthoryear{{Hynes} et~al.,}{{Hynes} et~al.}{2003}]{hynes03}
{Hynes} R.~I.,  et~al., 2003, \mn@doi [\mnras] {10.1046/j.1365-8711.2003.06938.x}, \href {https://ui.adsabs.harvard.edu/abs/2003MNRAS.345..292H} {345, 292}

\bibitem[\protect\citeauthoryear{{Ingram} \& {Motta}}{{Ingram} \& {Motta}}{2019}]{ingram19}
{Ingram} A.~R.,  {Motta} S.~E.,  2019, \mn@doi [\nar] {10.1016/j.newar.2020.101524}, \href {https://ui.adsabs.harvard.edu/abs/2019NewAR..8501524I} {85, 101524}

\bibitem[\protect\citeauthoryear{{Ingram}, {Done}  \& {Fragile}}{{Ingram} et~al.}{2009}]{ingram09}
{Ingram} A.,  {Done} C.,   {Fragile} P.~C.,  2009, \mn@doi [\mnras] {10.1111/j.1745-3933.2009.00693.x}, \href {https://ui.adsabs.harvard.edu/abs/2009MNRAS.397L.101I} {397, L101}

\bibitem[\protect\citeauthoryear{{Jim{\'e}nez-Ibarra}, {Mu{\~n}oz-Darias}, {Casares}, {Armas Padilla}  \& {Corral-Santana}}{{Jim{\'e}nez-Ibarra} et~al.}{2019}]{jimenez-ibarra19}
{Jim{\'e}nez-Ibarra} F.,  {Mu{\~n}oz-Darias} T.,  {Casares} J.,  {Armas Padilla} M.,   {Corral-Santana} J.~M.,  2019, \mn@doi [\mnras] {10.1093/mnras/stz2393}, \href {https://ui.adsabs.harvard.edu/abs/2019MNRAS.489.3420J} {489, 3420}

\bibitem[\protect\citeauthoryear{{Kawamuro} et~al.,}{{Kawamuro} et~al.}{2018}]{kawamuro18}
{Kawamuro} T.,  et~al., 2018, The Astronomer's Telegram, \href {https://ui.adsabs.harvard.edu/abs/2018ATel11399....1K} {11399, 1}

\bibitem[\protect\citeauthoryear{{Kimble} et~al.,}{{Kimble} et~al.}{1998}]{kimble98}
{Kimble} R.~A.,  et~al., 1998, \mn@doi [\apjl] {10.1086/311102}, \href {https://ui.adsabs.harvard.edu/abs/1998ApJ...492L..83K} {492, L83}

\bibitem[\protect\citeauthoryear{{King} \& {Ritter}}{{King} \& {Ritter}}{1998}]{king98}
{King} A.~R.,  {Ritter} H.,  1998, \mn@doi [\mnras] {10.1046/j.1365-8711.1998.01295.x}, \href {https://ui.adsabs.harvard.edu/abs/1998MNRAS.293L..42K} {293, L42}

\bibitem[\protect\citeauthoryear{{Koljonen}, {Long}, {Matthews}  \& {Knigge}}{{Koljonen} et~al.}{2023}]{koljonen23}
{Koljonen} K.~I.~I.,  {Long} K.~S.,  {Matthews} J.~H.,   {Knigge} C.,  2023, \mn@doi [\mnras] {10.1093/mnras/stad809}, \href {https://ui.adsabs.harvard.edu/abs/2023MNRAS.521.4190K} {521, 4190}

\bibitem[\protect\citeauthoryear{{Krimm} et~al.,}{{Krimm} et~al.}{2013}]{krimm13}
{Krimm} H.~A.,  et~al., 2013, \mn@doi [\apjs] {10.1088/0067-0049/209/1/14}, \href {https://ui.adsabs.harvard.edu/abs/2013ApJS..209...14K} {209, 14}

\bibitem[\protect\citeauthoryear{{Kumar}, {Dewangan}, {Singh}, {Gandhi}, {Papadakis}, {Tripathi}  \& {Mallick}}{{Kumar} et~al.}{2023}]{kumar2023}
{Kumar} S.,  {Dewangan} G.~C.,  {Singh} K.~P.,  {Gandhi} P.,  {Papadakis} I.~E.,  {Tripathi} P.,   {Mallick} L.,  2023, \mn@doi [\apj] {10.3847/1538-4357/acc941}, \href {https://ui.adsabs.harvard.edu/abs/2023ApJ...950...90K} {950, 90}

\bibitem[\protect\citeauthoryear{Lasota}{Lasota}{2001}]{lasota01}
Lasota J.,  2001, New Astronomy Reviews, 45, 449

\bibitem[\protect\citeauthoryear{{Liszt}}{{Liszt}}{2014}]{liszt14}
{Liszt} H.,  2014, \mn@doi [\apj] {10.1088/0004-637X/780/1/10}, \href {https://ui.adsabs.harvard.edu/abs/2014ApJ...780...10L} {780, 10}

\bibitem[\protect\citeauthoryear{{Ma} et~al.,}{{Ma} et~al.}{2021}]{ma21}
{Ma} X.,  et~al., 2021, \mn@doi [Nature Astronomy] {10.1038/s41550-020-1192-2}, \href {https://ui.adsabs.harvard.edu/abs/2021NatAs...5...94M} {5, 94}

\bibitem[\protect\citeauthoryear{{Mao}, {Yu}, {Zhang}, {Yan}, {Rapisarda}, {Wang}  \& {Bai}}{{Mao} et~al.}{2022}]{mao22}
{Mao} D.-M.,  {Yu} W.-F.,  {Zhang} J.-J.,  {Yan} Z.,  {Rapisarda} S.,  {Wang} X.-F.,   {Bai} J.-M.,  2022, \mn@doi [Research in Astronomy and Astrophysics] {10.1088/1674-4527/ac538a}, \href {https://ui.adsabs.harvard.edu/abs/2022RAA....22d5009M} {22, 045009}

\bibitem[\protect\citeauthoryear{{Markoff}, {Falcke}  \& {Fender}}{{Markoff} et~al.}{2001}]{markoff01}
{Markoff} S.,  {Falcke} H.,   {Fender} R.,  2001, \mn@doi [\aap] {10.1051/0004-6361:20010420}, \href {https://ui.adsabs.harvard.edu/abs/2001A&A...372L..25M} {372, L25}

\bibitem[\protect\citeauthoryear{{Mata S{\'a}nchez}, {Rau}, {{\'A}lvarez Hern{\'a}ndez}, {van Grunsven}, {Torres}  \& {Jonker}}{{Mata S{\'a}nchez} et~al.}{2021}]{matasanchez21}
{Mata S{\'a}nchez} D.,  {Rau} A.,  {{\'A}lvarez Hern{\'a}ndez} A.,  {van Grunsven} T.~F.~J.,  {Torres} M.~A.~P.,   {Jonker} P.~G.,  2021, \mn@doi [\mnras] {10.1093/mnras/stab1714}, \href {https://ui.adsabs.harvard.edu/abs/2021MNRAS.506..581M} {506, 581}

\bibitem[\protect\citeauthoryear{{Mauche}, {Lee}  \& {Kallman}}{{Mauche} et~al.}{1997}]{mauche97}
{Mauche} C.~W.,  {Lee} Y.~P.,   {Kallman} T.~R.,  1997, \mn@doi [\apj] {10.1086/303717}, \href {https://ui.adsabs.harvard.edu/abs/1997ApJ...477..832M} {477, 832}

\bibitem[\protect\citeauthoryear{{Miko{\l}ajewska}, {Zdziarski}, {Zi{\'o}{\l}kowski}, {Torres}  \& {Casares}}{{Miko{\l}ajewska} et~al.}{2022}]{mikolajewska22}
{Miko{\l}ajewska} J.,  {Zdziarski} A.~A.,  {Zi{\'o}{\l}kowski} J.,  {Torres} M. A.~P.,   {Casares} J.,  2022, \mn@doi [\apj] {10.3847/1538-4357/ac6099}, \href {https://ui.adsabs.harvard.edu/abs/2022ApJ...930....9M} {930, 9}

\bibitem[\protect\citeauthoryear{{Miller}, {Raymond}, {Fabian}, {Steeghs}, {Homan}, {Reynolds}, {van der Klis}  \& {Wijnands}}{{Miller} et~al.}{2006a}]{miller06a}
{Miller} J.~M.,  {Raymond} J.,  {Fabian} A.,  {Steeghs} D.,  {Homan} J.,  {Reynolds} C.,  {van der Klis} M.,   {Wijnands} R.,  2006a, \mn@doi [\nat] {10.1038/nature04912}, \href {https://ui.adsabs.harvard.edu/abs/2006Natur.441..953M} {441, 953}

\bibitem[\protect\citeauthoryear{{Miller} et~al.,}{{Miller} et~al.}{2006b}]{miller06b}
{Miller} J.~M.,  et~al., 2006b, \mn@doi [\apj] {10.1086/504673}, \href {https://ui.adsabs.harvard.edu/abs/2006ApJ...646..394M} {646, 394}

\bibitem[\protect\citeauthoryear{{Miller}, {Raymond}, {Reynolds}, {Fabian}, {Kallman}  \& {Homan}}{{Miller} et~al.}{2008}]{miller08}
{Miller} J.~M.,  {Raymond} J.,  {Reynolds} C.~S.,  {Fabian} A.~C.,  {Kallman} T.~R.,   {Homan} J.,  2008, \mn@doi [\apj] {10.1086/588521}, \href {https://ui.adsabs.harvard.edu/abs/2008ApJ...680.1359M} {680, 1359}

\bibitem[\protect\citeauthoryear{{Motch}, {Ilovaisky}  \& {Chevalier}}{{Motch} et~al.}{1982}]{motch82}
{Motch} C.,  {Ilovaisky} S.~A.,   {Chevalier} C.,  1982, \aap, \href {https://ui.adsabs.harvard.edu/abs/1982A&A...109L...1M} {109, L1}

\bibitem[\protect\citeauthoryear{{Motch}, {Ricketts}, {Page}, {Ilovaisky}  \& {Chevalier}}{{Motch} et~al.}{1983}]{motch83}
{Motch} C.,  {Ricketts} M.~J.,  {Page} C.~G.,  {Ilovaisky} S.~A.,   {Chevalier} C.,  1983, \aap, \href {https://ui.adsabs.harvard.edu/abs/1983A&A...119..171M} {119, 171}

\bibitem[\protect\citeauthoryear{{Motta}}{{Motta}}{2016}]{motta16}
{Motta} S.~E.,  2016, \mn@doi [Astronomische Nachrichten] {10.1002/asna.201612320}, \href {https://ui.adsabs.harvard.edu/abs/2016AN....337..398M} {337, 398}

\bibitem[\protect\citeauthoryear{{Mu{\~n}oz-Darias} \& {Ponti}}{{Mu{\~n}oz-Darias} \& {Ponti}}{2022}]{munoz-darias22}
{Mu{\~n}oz-Darias} T.,  {Ponti} G.,  2022, \mn@doi [\aap] {10.1051/0004-6361/202243769}, \href {https://ui.adsabs.harvard.edu/abs/2022A&A...664A.104M} {664, A104}

\bibitem[\protect\citeauthoryear{{Mu{\~n}oz-Darias}, {Motta}  \& {Belloni}}{{Mu{\~n}oz-Darias} et~al.}{2011}]{munoz-darias11}
{Mu{\~n}oz-Darias} T.,  {Motta} S.,   {Belloni} T.~M.,  2011, \mn@doi [\mnras] {10.1111/j.1365-2966.2010.17476.x}, \href {https://ui.adsabs.harvard.edu/abs/2011MNRAS.410..679M} {410, 679}

\bibitem[\protect\citeauthoryear{{Mu{\~n}oz-Darias} et~al.,}{{Mu{\~n}oz-Darias} et~al.}{2016}]{munoz-darias16}
{Mu{\~n}oz-Darias} T.,  et~al., 2016, \mn@doi [\nat] {10.1038/nature17446}, \href {https://ui.adsabs.harvard.edu/abs/2016Natur.534...75M} {534, 75}

\bibitem[\protect\citeauthoryear{{Mu{\~n}oz-Darias}, {Torres}  \& {Garcia}}{{Mu{\~n}oz-Darias} et~al.}{2018}]{munoz-darias18}
{Mu{\~n}oz-Darias} T.,  {Torres} M. A.~P.,   {Garcia} M.~R.,  2018, \mn@doi [\mnras] {10.1093/mnras/sty1711}, \href {https://ui.adsabs.harvard.edu/abs/2018MNRAS.479.3987M} {479, 3987}

\bibitem[\protect\citeauthoryear{{Mu{\~n}oz-Darias} et~al.,}{{Mu{\~n}oz-Darias} et~al.}{2019}]{munoz-darias19}
{Mu{\~n}oz-Darias} T.,  et~al., 2019, \apjl, \href {https://ui.adsabs.harvard.edu/abs/2019ApJ...879L...4M} {879, L4}

\bibitem[\protect\citeauthoryear{{Mudambi}, {Maqbool}, {Misra}, {Hebbar}, {Yadav}, {Gudennavar}  \& {S.~G.}}{{Mudambi} et~al.}{2020}]{mudambi20}
{Mudambi} S.~P.,  {Maqbool} B.,  {Misra} R.,  {Hebbar} S.,  {Yadav} J.~S.,  {Gudennavar} S.~B.,   {S.~G.} B.,  2020, \mn@doi [\apjl] {10.3847/2041-8213/ab66bc}, \href {https://ui.adsabs.harvard.edu/abs/2020ApJ...889L..17M} {889, L17}

\bibitem[\protect\citeauthoryear{{Neilsen} \& {Lee}}{{Neilsen} \& {Lee}}{2009}]{neilsen09}
{Neilsen} J.,  {Lee} J.~C.,  2009, \mn@doi [\nat] {10.1038/nature07680}, \href {https://ui.adsabs.harvard.edu/abs/2009Natur.458..481N} {458, 481}

\bibitem[\protect\citeauthoryear{{Novikov} \& {Thorne}}{{Novikov} \& {Thorne}}{1973}]{novikov73}
{Novikov} I.~D.,  {Thorne} K.~S.,  1973, in Black Holes (Les Astres Occlus). pp 343--450

\bibitem[\protect\citeauthoryear{{{\"O}zbey Arabac{\i}}, {Kalemci}, {Din{\c{c}}er}, {Bailyn}, {Altamirano}  \& {Ak}}{{{\"O}zbey Arabac{\i}} et~al.}{2022}]{arabaci22}
{{\"O}zbey Arabac{\i}} M.,  {Kalemci} E.,  {Din{\c{c}}er} T.,  {Bailyn} C.~D.,  {Altamirano} D.,   {Ak} T.,  2022, \mn@doi [\mnras] {10.1093/mnras/stac1574}, \href {https://ui.adsabs.harvard.edu/abs/2022MNRAS.514.3894O} {514, 3894}

\bibitem[\protect\citeauthoryear{{Paice} et~al.,}{{Paice} et~al.}{2021}]{paice21}
{Paice} J.~A.,  et~al., 2021, \mn@doi [\mnras] {10.1093/mnras/stab1531}, \href {https://ui.adsabs.harvard.edu/abs/2021MNRAS.505.3452P} {505, 3452}

\bibitem[\protect\citeauthoryear{{Ponti}, {Fender}, {Begelman}, {Dunn}, {Neilsen}  \& {Coriat}}{{Ponti} et~al.}{2012}]{ponti12}
{Ponti} G.,  {Fender} R.~P.,  {Begelman} M.~C.,  {Dunn} R.~J.~H.,  {Neilsen} J.,   {Coriat} M.,  2012, \mn@doi [\mnras] {10.1111/j.1745-3933.2012.01224.x}, \href {https://ui.adsabs.harvard.edu/abs/2012MNRAS.422L..11P} {422, L11}

\bibitem[\protect\citeauthoryear{{Ponti}, {Mu{\~n}oz-Darias}  \& {Fender}}{{Ponti} et~al.}{2014}]{ponti14}
{Ponti} G.,  {Mu{\~n}oz-Darias} T.,   {Fender} R.~P.,  2014, \mn@doi [\mnras] {10.1093/mnras/stu1742}, \href {https://ui.adsabs.harvard.edu/abs/2014MNRAS.444.1829P} {444, 1829}

\bibitem[\protect\citeauthoryear{{Ponti}, {Bianchi}, {Mu{\~n}oz-Darias}, {De}, {Fender}  \& {Merloni}}{{Ponti} et~al.}{2016}]{ponti16}
{Ponti} G.,  {Bianchi} S.,  {Mu{\~n}oz-Darias} T.,  {De} K.,  {Fender} R.,   {Merloni} A.,  2016, \mn@doi [Astronomische Nachrichten] {10.1002/asna.201612339}, \href {https://ui.adsabs.harvard.edu/abs/2016AN....337..512P} {337, 512}

\bibitem[\protect\citeauthoryear{Postma \& Leahy}{Postma \& Leahy}{2017}]{postma2017}
Postma J.~E.,  Leahy D.,  2017, \mn@doi [Publications of the Astronomical Society of the Pacific] {10.1088/1538-3873/aa8800}, 129, 115002

\bibitem[\protect\citeauthoryear{Remillard \& McClintock}{Remillard \& McClintock}{2006}]{remillard06}
Remillard R.~A.,  McClintock J.~E.,  2006, Annual Review of Astronomy and Astrophysics, 44, 49

\bibitem[\protect\citeauthoryear{{Russell}, {Miller-Jones}, {Maccarone}, {Yang}, {Fender}  \& {Lewis}}{{Russell} et~al.}{2011}]{russell11}
{Russell} D.~M.,  {Miller-Jones} J.~C.~A.,  {Maccarone} T.~J.,  {Yang} Y.~J.,  {Fender} R.~P.,   {Lewis} F.,  2011, \mn@doi [\apjl] {10.1088/2041-8205/739/1/L19}, \href {https://ui.adsabs.harvard.edu/abs/2011ApJ...739L..19R} {739, L19}

\bibitem[\protect\citeauthoryear{{S{\'a}nchez-Sierras} \& {Mu{\~n}oz-Darias}}{{S{\'a}nchez-Sierras} \& {Mu{\~n}oz-Darias}}{2020}]{sanchez-sierras20}
{S{\'a}nchez-Sierras} J.,  {Mu{\~n}oz-Darias} T.,  2020, \mn@doi [\aap] {10.1051/0004-6361/202038406}, \href {https://ui.adsabs.harvard.edu/abs/2020A&A...640L...3S} {640, L3}

\bibitem[\protect\citeauthoryear{{Schenker}, {King}, {Kolb}, {Wynn}  \& {Zhang}}{{Schenker} et~al.}{2002}]{schenker02}
{Schenker} K.,  {King} A.~R.,  {Kolb} U.,  {Wynn} G.~A.,   {Zhang} Z.,  2002, \mn@doi [\mnras] {10.1046/j.1365-8711.2002.05999.x}, \href {https://ui.adsabs.harvard.edu/abs/2002MNRAS.337.1105S} {337, 1105}

\bibitem[\protect\citeauthoryear{{Schlafly} \& {Finkbeiner}}{{Schlafly} \& {Finkbeiner}}{2011}]{Schlafly11}
{Schlafly} E.~F.,  {Finkbeiner} D.~P.,  2011, \mn@doi [\apj] {10.1088/0004-637X/737/2/103}, \href {https://ui.adsabs.harvard.edu/abs/2011ApJ...737..103S} {737, 103}

\bibitem[\protect\citeauthoryear{{Schnittman}, {Homan}  \& {Miller}}{{Schnittman} et~al.}{2006}]{schnittman06}
{Schnittman} J.~D.,  {Homan} J.,   {Miller} J.~M.,  2006, \mn@doi [\apj] {10.1086/500923}, \href {https://ui.adsabs.harvard.edu/abs/2006ApJ...642..420S} {642, 420}

\bibitem[\protect\citeauthoryear{Shakura \& Sunyaev}{Shakura \& Sunyaev}{1973}]{shakura73}
Shakura N.,  Sunyaev R.,  1973, \aap, 24, 337

\bibitem[\protect\citeauthoryear{{Shimura} \& {Takahara}}{{Shimura} \& {Takahara}}{1995}]{shimura95}
{Shimura} T.,  {Takahara} F.,  1995, \mn@doi [\apj] {10.1086/175300}, \href {https://ui.adsabs.harvard.edu/abs/1995ApJ...440..610S} {440, 610}

\bibitem[\protect\citeauthoryear{{Stella} \& {Vietri}}{{Stella} \& {Vietri}}{1998}]{stella98}
{Stella} L.,  {Vietri} M.,  1998, \mn@doi [\apjl] {10.1086/311075}, \href {https://ui.adsabs.harvard.edu/abs/1998ApJ...492L..59S} {492, L59}

\bibitem[\protect\citeauthoryear{{Stella}, {Vietri}  \& {Morsink}}{{Stella} et~al.}{1999}]{stella99}
{Stella} L.,  {Vietri} M.,   {Morsink} S.~M.,  1999, \mn@doi [\apjl] {10.1086/312291}, \href {https://ui.adsabs.harvard.edu/abs/1999ApJ...524L..63S} {524, L63}

\bibitem[\protect\citeauthoryear{{Stiele} \& {Kong}}{{Stiele} \& {Kong}}{2020}]{stiele2020}
{Stiele} H.,  {Kong} A.~K.~H.,  2020, \mn@doi [\apj] {10.3847/1538-4357/ab64ef}, \href {https://ui.adsabs.harvard.edu/abs/2020ApJ...889..142S} {889, 142}

\bibitem[\protect\citeauthoryear{{Summons}, {Ar{\'e}valo}, {McHardy}, {Uttley}  \& {Bhaskar}}{{Summons} et~al.}{2007}]{summons07}
{Summons} D.~P.,  {Ar{\'e}valo} P.,  {McHardy} I.~M.,  {Uttley} P.,   {Bhaskar} A.,  2007, \mn@doi [\mnras] {10.1111/j.1365-2966.2006.11797.x}, \href {https://ui.adsabs.harvard.edu/abs/2007MNRAS.378..649S} {378, 649}

\bibitem[\protect\citeauthoryear{{Tagger} \& {Pellat}}{{Tagger} \& {Pellat}}{1999}]{tagger99}
{Tagger} M.,  {Pellat} R.,  1999, \mn@doi [\aap] {10.48550/arXiv.astro-ph/9907267}, \href {https://ui.adsabs.harvard.edu/abs/1999A&A...349.1003T} {349, 1003}

\bibitem[\protect\citeauthoryear{Tandon et~al.,}{Tandon et~al.}{2017}]{tandon2017orbit}
Tandon S.,  et~al., 2017, The Astronomical Journal, 154, 128

\bibitem[\protect\citeauthoryear{Tandon et~al.,}{Tandon et~al.}{2020}]{tandon2020additional}
Tandon S.,  et~al., 2020, The Astronomical Journal, 159, 158

\bibitem[\protect\citeauthoryear{{Tetarenko}, {Dubus}, {Lasota}, {Heinke}  \& {Sivakoff}}{{Tetarenko} et~al.}{2018a}]{tetarenko18b}
{Tetarenko} B.~E.,  {Dubus} G.,  {Lasota} J.~P.,  {Heinke} C.~O.,   {Sivakoff} G.~R.,  2018a, \mn@doi [\mnras] {10.1093/mnras/sty1798}, \href {https://ui.adsabs.harvard.edu/abs/2018MNRAS.480....2T} {480, 2}

\bibitem[\protect\citeauthoryear{{Tetarenko}, {Petitpas}, {Sivakoff}, {Miller-Jones}, {Russell}, {Schieven}  \& {Jacpot Xrb Collaboration}}{{Tetarenko} et~al.}{2018b}]{tetarenko18a}
{Tetarenko} A.~J.,  {Petitpas} G.,  {Sivakoff} G.~R.,  {Miller-Jones} J.~C.~A.,  {Russell} T.~D.,  {Schieven} G.,   {Jacpot Xrb Collaboration} 2018b, The Astronomer's Telegram, \href {https://ui.adsabs.harvard.edu/abs/2018ATel11831....1T} {11831, 1}

\bibitem[\protect\citeauthoryear{{Tetarenko}, {Shaw}, {Manrow}, {Charles}, {Miller}, {Russell}  \& {Tetarenko}}{{Tetarenko} et~al.}{2021a}]{tetarenko21b}
{Tetarenko} B.~E.,  {Shaw} A.~W.,  {Manrow} E.~R.,  {Charles} P.~A.,  {Miller} J.~M.,  {Russell} T.~D.,   {Tetarenko} A.~J.,  2021a, \mn@doi [\mnras] {10.1093/mnras/staa3861}, \href {https://ui.adsabs.harvard.edu/abs/2021MNRAS.501.3406T} {501, 3406}

\bibitem[\protect\citeauthoryear{{Tetarenko} et~al.,}{{Tetarenko} et~al.}{2021b}]{tetarenko21a}
{Tetarenko} A.~J.,  et~al., 2021b, \mn@doi [\mnras] {10.1093/mnras/stab820}, \href {https://ui.adsabs.harvard.edu/abs/2021MNRAS.504.3862T} {504, 3862}

\bibitem[\protect\citeauthoryear{{Thomas} et~al.,}{{Thomas} et~al.}{2022}]{thomas22}
{Thomas} J.~K.,  et~al., 2022, \mn@doi [\mnras] {10.1093/mnrasl/slab132}, \href {https://ui.adsabs.harvard.edu/abs/2022MNRAS.513L..35T} {513, L35}

\bibitem[\protect\citeauthoryear{{Titarchuk} \& {Fiorito}}{{Titarchuk} \& {Fiorito}}{2004}]{titarchuk04}
{Titarchuk} L.,  {Fiorito} R.,  2004, \mn@doi [\apj] {10.1086/422573}, \href {https://ui.adsabs.harvard.edu/abs/2004ApJ...612..988T} {612, 988}

\bibitem[\protect\citeauthoryear{{Torres}, {Casares}, {Jim{\'e}nez-Ibarra}, {{\'A}lvarez-Hern{\'a}ndez}, {Mu{\~n}oz-Darias}, {Armas Padilla}, {Jonker}  \& {Heida}}{{Torres} et~al.}{2020}]{torres20}
{Torres} M.~A.~P.,  {Casares} J.,  {Jim{\'e}nez-Ibarra} F.,  {{\'A}lvarez-Hern{\'a}ndez} A.,  {Mu{\~n}oz-Darias} T.,  {Armas Padilla} M.,  {Jonker} P.~G.,   {Heida} M.,  2020, \mn@doi [\apjl] {10.3847/2041-8213/ab863a}, \href {https://ui.adsabs.harvard.edu/abs/2020ApJ...893L..37T} {893, L37}

\bibitem[\protect\citeauthoryear{{Tucker} et~al.,}{{Tucker} et~al.}{2018}]{tucker18}
{Tucker} M.~A.,  et~al., 2018, \mn@doi [\apjl] {10.3847/2041-8213/aae88a}, \href {https://ui.adsabs.harvard.edu/abs/2018ApJ...867L...9T} {867, L9}

\bibitem[\protect\citeauthoryear{{Varni{\`e}re}, {Rodriguez}  \& {Tagger}}{{Varni{\`e}re} et~al.}{2002}]{varniere02}
{Varni{\`e}re} P.,  {Rodriguez} J.,   {Tagger} M.,  2002, \mn@doi [\aap] {10.1051/0004-6361:20020401}, \href {https://ui.adsabs.harvard.edu/abs/2002A&A...387..497V} {387, 497}

\bibitem[\protect\citeauthoryear{{Varni{\`e}re}, {Tagger}  \& {Rodriguez}}{{Varni{\`e}re} et~al.}{2012}]{varniere12}
{Varni{\`e}re} P.,  {Tagger} M.,   {Rodriguez} J.,  2012, \mn@doi [\aap] {10.1051/0004-6361/201116698}, \href {https://ui.adsabs.harvard.edu/abs/2012A&A...545A..40V} {545, A40}

\bibitem[\protect\citeauthoryear{{Vaughan}}{{Vaughan}}{2005}]{vaughan05}
{Vaughan} S.,  2005, \mn@doi [\aap] {10.1051/0004-6361:20041453}, \href {https://ui.adsabs.harvard.edu/abs/2005A&A...431..391V} {431, 391}

\bibitem[\protect\citeauthoryear{{Vaughan}, {Edelson}, {Warwick}  \& {Uttley}}{{Vaughan} et~al.}{2003}]{vaughan03}
{Vaughan} S.,  {Edelson} R.,  {Warwick} R.~S.,   {Uttley} P.,  2003, \mn@doi [\mnras] {10.1046/j.1365-2966.2003.07042.x}, \href {https://ui.adsabs.harvard.edu/abs/2003MNRAS.345.1271V} {345, 1271}

\bibitem[\protect\citeauthoryear{{Veledina} \& {Poutanen}}{{Veledina} \& {Poutanen}}{2015}]{veledina15}
{Veledina} A.,  {Poutanen} J.,  2015, \mn@doi [\mnras] {10.1093/mnras/stu2737}, \href {https://ui.adsabs.harvard.edu/abs/2015MNRAS.448..939V} {448, 939}

\bibitem[\protect\citeauthoryear{{Veledina}, {Poutanen}  \& {Vurm}}{{Veledina} et~al.}{2013}]{veledina13}
{Veledina} A.,  {Poutanen} J.,   {Vurm} I.,  2013, \mn@doi [\mnras] {10.1093/mnras/stt124}, \href {https://ui.adsabs.harvard.edu/abs/2013MNRAS.430.3196V} {430, 3196}

\bibitem[\protect\citeauthoryear{Virtanen et~al.,}{Virtanen et~al.}{2020}]{scipy2020}
Virtanen P.,  et~al., 2020, \mn@doi [Nature Methods] {10.1038/s41592-019-0686-2}, \href {https://rdcu.be/b08Wh} {17, 261}

\bibitem[\protect\citeauthoryear{{Woodgate} et~al.,}{{Woodgate} et~al.}{1998}]{woodgate98}
{Woodgate} B.~E.,  et~al., 1998, \mn@doi [\pasp] {10.1086/316243}, \href {https://ui.adsabs.harvard.edu/abs/1998PASP..110.1183W} {110, 1183}

\bibitem[\protect\citeauthoryear{{Yoshitake} et~al.,}{{Yoshitake} et~al.}{2022}]{yoshitake22}
{Yoshitake} T.,  et~al., 2022, \mn@doi [\pasj] {10.1093/pasj/psac038}, \href {https://ui.adsabs.harvard.edu/abs/2022PASJ...74..805Y} {74, 805}

\bibitem[\protect\citeauthoryear{{Yu}, {Zhang}, {Yan}, {Wang}  \& {Bai}}{{Yu} et~al.}{2018}]{yu2018a}
{Yu} W.,  {Zhang} J.,  {Yan} Z.,  {Wang} X.,   {Bai} J.,  2018, The Astronomer's Telegram, \href {https://ui.adsabs.harvard.edu/abs/2018ATel11510....1Y} {11510, 1}

\bibitem[\protect\citeauthoryear{{Zdziarski} \& {Gierli{\'n}ski}}{{Zdziarski} \& {Gierli{\'n}ski}}{2004}]{zdziarski04}
{Zdziarski} A.~A.,  {Gierli{\'n}ski} M.,  2004, \mn@doi [Progress of Theoretical Physics Supplement] {10.1143/PTPS.155.99}, \href {https://ui.adsabs.harvard.edu/abs/2004PThPS.155...99Z} {155, 99}

\bibitem[\protect\citeauthoryear{{Zhao} et~al.,}{{Zhao} et~al.}{2021}]{zhao21}
{Zhao} X.,  et~al., 2021, \mn@doi [\apj] {10.3847/1538-4357/ac07a9}, \href {https://ui.adsabs.harvard.edu/abs/2021ApJ...916..108Z} {916, 108}

\bibitem[\protect\citeauthoryear{{della Valle}, {Benetti}, {Cappellaro}  \& {Wheeler}}{{della Valle} et~al.}{1997}]{DellaValle97}
{della Valle} M.,  {Benetti} S.,  {Cappellaro} E.,   {Wheeler} C.,  1997, \aap, \href {https://ui.adsabs.harvard.edu/abs/1997A&A...318..179D} {318, 179}

\bibitem[\protect\citeauthoryear{{van Paradijs}}{{van Paradijs}}{1996}]{vanparadijs96}
{van Paradijs} J.,  1996, \mn@doi [\apjl] {10.1086/310100}, \href {https://ui.adsabs.harvard.edu/abs/1996ApJ...464L.139V} {464, L139}

\bibitem[\protect\citeauthoryear{{van Paradijs} \& {McClintock}}{{van Paradijs} \& {McClintock}}{1994}]{vanparadijs94}
{van Paradijs} J.,  {McClintock} J.~E.,  1994, \aap, \href {https://ui.adsabs.harvard.edu/abs/1994A&A...290..133V} {290, 133}

\bibitem[\protect\citeauthoryear{{van der Klis}}{{van der Klis}}{1989}]{vanderklis89}
{van der Klis} M.,  1989, in {{\"O}gelman} H.,  {van den Heuvel} E.~P.~J.,  eds,  NATO Advanced Study Institute (ASI) Series C Vol. 262, Timing Neutron Stars. p.~27

\bibitem[\protect\citeauthoryear{{van der Klis}}{{van der Klis}}{1997}]{vanderklis97}
{van der Klis} M.,  1997, in {Maoz} D.,  {Sternberg} A.,   {Leibowitz} E.~M.,  eds,  Astrophysics and Space Science Library Vol. 218, Astronomical Time Series. p.~121

\bibitem[\protect\citeauthoryear{{van der Walt}, {Colbert}  \& {Varoquaux}}{{van der Walt} et~al.}{2011}]{vanderwalt11}
{van der Walt} S.,  {Colbert} S.~C.,   {Varoquaux} G.,  2011, \mn@doi [Computing in Science and Engineering] {10.1109/MCSE.2011.37}, \href {https://ui.adsabs.harvard.edu/abs/2011CSE....13b..22V} {13, 22}

\makeatother
\end{thebibliography}







\bsp	
\label{lastpage}
\end{document}